\documentclass[preprint,12pt]{elsarticle}

\usepackage{amssymb}
\usepackage{amsmath}
\usepackage{natbib}

\usepackage{url}
\newcommand{\norm}[1]{\left\lVert#1\right\rVert}
\DeclareMathOperator*{\argmin}{argmin}
\newcommand*{\argminl}{\argmin\limits}

\begin{document}

\begin{frontmatter}



\title{A System for the Generation of Synthetic Wide Area Aerial Surveillance Imagery}

\author[uol]{Elias J Griffith}
 \ead{ejg@liverpool.ac.uk}
 \author[uol]{Chinmaya Mishra}
 \ead{cmishra@liverpool.ac.uk}
\author[uol]{Jason F. Ralph\corref{cor1}}
 \ead{jfralph@liverpool.ac.uk}
\author[uol]{Simon Maskell}
 \ead{s.maskell@liverpool.ac.uk}

 \cortext[cor1]{Corresponding author}

\address[uol]{Department of Electrical Engineering and Electronics, University of Liverpool,  Brownlow Hill, Liverpool, L69 3GJ, UK.}

\begin{abstract}
The development, benchmarking and validation of aerial Persistent Surveillance (PS) algorithms requires access to specialist Wide Area Aerial Surveillance (WAAS) datasets. Such datasets are difficult to obtain and are often extremely large both in spatial resolution and temporal duration. This paper outlines an approach to the simulation of complex urban environments and demonstrates the viability of using this approach for the generation of simulated sensor data, corresponding to the use of wide area imaging systems for surveillance and reconnaissance applications. This provides a cost-effective method to generate datasets for vehicle tracking algorithms and anomaly detection methods. The system fuses the Simulation of Urban Mobility (SUMO) traffic simulator with a MATLAB controller and an image generator to create scenes containing uninterrupted door-to-door journeys across large areas of the urban environment. This `pattern-of-life' approach provides three-dimensional visual information with natural movement and traffic flows. This can then be used to provide simulated sensor measurements (e.g. visual band and infrared video imagery) and automatic access to ground-truth data for the evaluation of multi-target tracking systems.   
\end{abstract}

\begin{keyword}
Persistent Surveillance,  Image Generation, City Simulation, WAAS, WAMI.
\end{keyword}

\end{frontmatter}

{\section{Introduction}\label{fig:chapter1}}

Wide Area Persistent Surveillance is the use of a sensor or sensor network to monitor a very large area, continuously and over long periods of time~\cite{menthe2012future}. This surveillance can provide a variety of information; to allow a forensic analysis of a significant incident (good or bad), or a general analysis of the pattern of life in order to optimize the flow of people or other traffic through a city~\cite{neirotti2014current}. In principle, given continuous coverage, it can allow persons of interest to be tracked from one location to the next~\cite{menthe2012future,levchuk2013learning} or track stolen vehicles traveling through a road network~\cite{collins2000system}.  Although some data could be provided by ground-based sensors, such as Closed Circuit Television (CCTV) cameras, the simplest way to provide continuous coverage of a city-sized area is to use airborne sensors, such as a Wide Area Aerial Surveillance (WAAS) system~\cite{menthe2012future} incorporating a very high resolution, wide field of view camera system; capable of producing images of the order of a hundred megapixels~\cite{cohenour2015camera} to several gigapixels~\cite{argus_datasheet} in size and at a relatively high frame rate. 

The data produced by these devices is of considerable interest to Big Data analysts. The data rate of the sensor will often far exceed the capacity of the communications networks used to transfer the data to the ground for processing~\cite{baraniuk2011more}, and it can easily exceed the ability and availability of human analysts to interpret the data once it has been transferred~\cite{menthe2012future}. This is particularly true when the area being surveiled is a complex urban environment, with different types of road traffic (local traffic, arterial flow, and public transport) and a huge variety of behavior exhibited by the individuals. This means that useful or relevant information normally needs to be extracted from the raw sensor imagery automatically, before being forwarded to the next level of processing. The development of automated systems to do this data selection and prioritization is made difficult by a lack of available data on which to optimize the processing. Public datasets monitoring different activities across very large urban areas are not readily available, and benchmark data are not as common as in some other standard image processing applications. Generating the data requires specialized and expensive sensors (which may be restricted to military users), and the use of co-ordinated flights over heavily populated, urban areas. Even when these issues have been resolved, recording large amounts of surveillance data over an city could give rise to legal concerns over the privacy of the individuals being monitored. It can be difficult to anonymize such data, because people and vehicles may be traced to known housing or other locations. Once recorded, there are further problems associated with distributing the large quantities of data involved and with the security of the final data set. In spite of these difficulties, some relevant data sets do exist. Examples of existing WAAS datasets include the Columbia Large Image Format (CLIF 2006, CLIF 2007), the Greene Town Centre (Greene 07), the Wright Patterson Air Force Base (WPAFB 2009), and the Minor Area Motion Imagery (MAMI 2013) datasets, all available from the SDMS website~\cite{sdms_website}. 

The data sets that are available for academic study are important for the validation of automated WAAS data processing methods, but they will never offer the flexibility required for the design and optimization of specialist tools and rapid prototyping of new algorithms. In these areas, synthetic image generation has several advantages over pre-recorded datasets, and research into WAAS sensor development could benefit hugely from developments in modelling and simulation methods. In addition to solving many of the problems described in the previous paragraph, synthetic data allows an exploration of unusual camera angles, configurations and parameters, the insertion of special behaviors and events, and (importantly) access to the exact truth data used to generate the imagery. The simulation of wide area video sequences has specific requirements: (i) Consistency -- the same quality and level of detail must exist across the simulated area and throughout the video; (ii) Coherency -- interactions between people and between vehicles can have far reaching influence and extend throughout the city area; and (iii) Big Data -- smaller simulations can choose between simulating a smaller area or duration, whereas wide area persistent surveillance requires both large areas and long periods of time. 

\begin{figure}
\centering
\includegraphics[width=0.8\columnwidth]{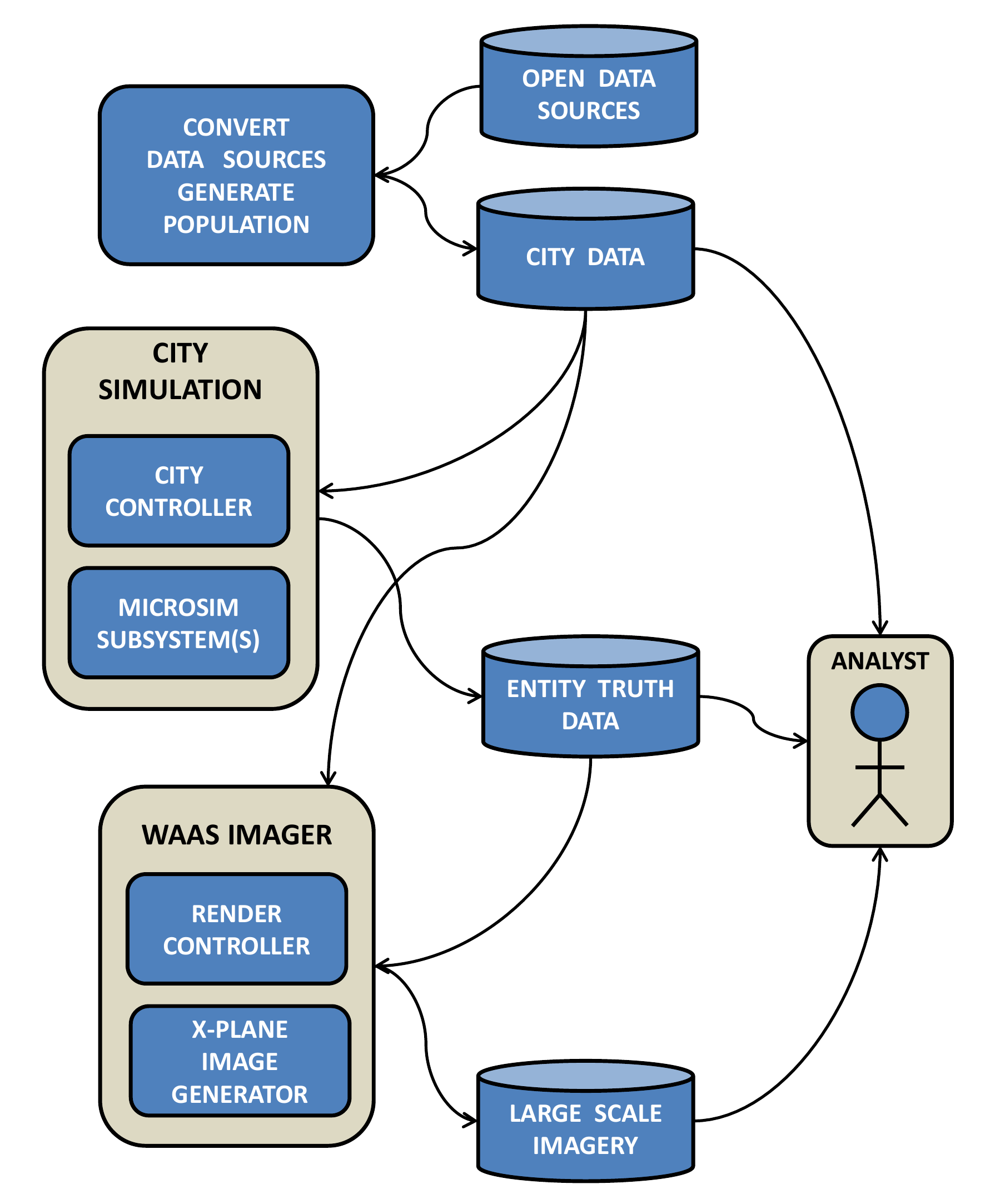}
\caption{System components for the Wide Area Simulator.  Open source data is drawn from the Open Street Map project (OSM), and the United States Geological Survey (USGS).  These data sources are converted to useable formats for the City Simulator and image generator. The City Simulator contains a number of expandable subsystems controled by a City Controller which outputs positional state data. The WAAS Imager uses the positional data to generate video and still images from a MATLAB controled image generator (X-Plane 10).  The positional state data is the truth data for the imagery, and can be processed by other analysis and verification tools.}
\label{fig:full_system_framework}
\end{figure}

\begin{figure}
\centering
\includegraphics[width=0.8\columnwidth]{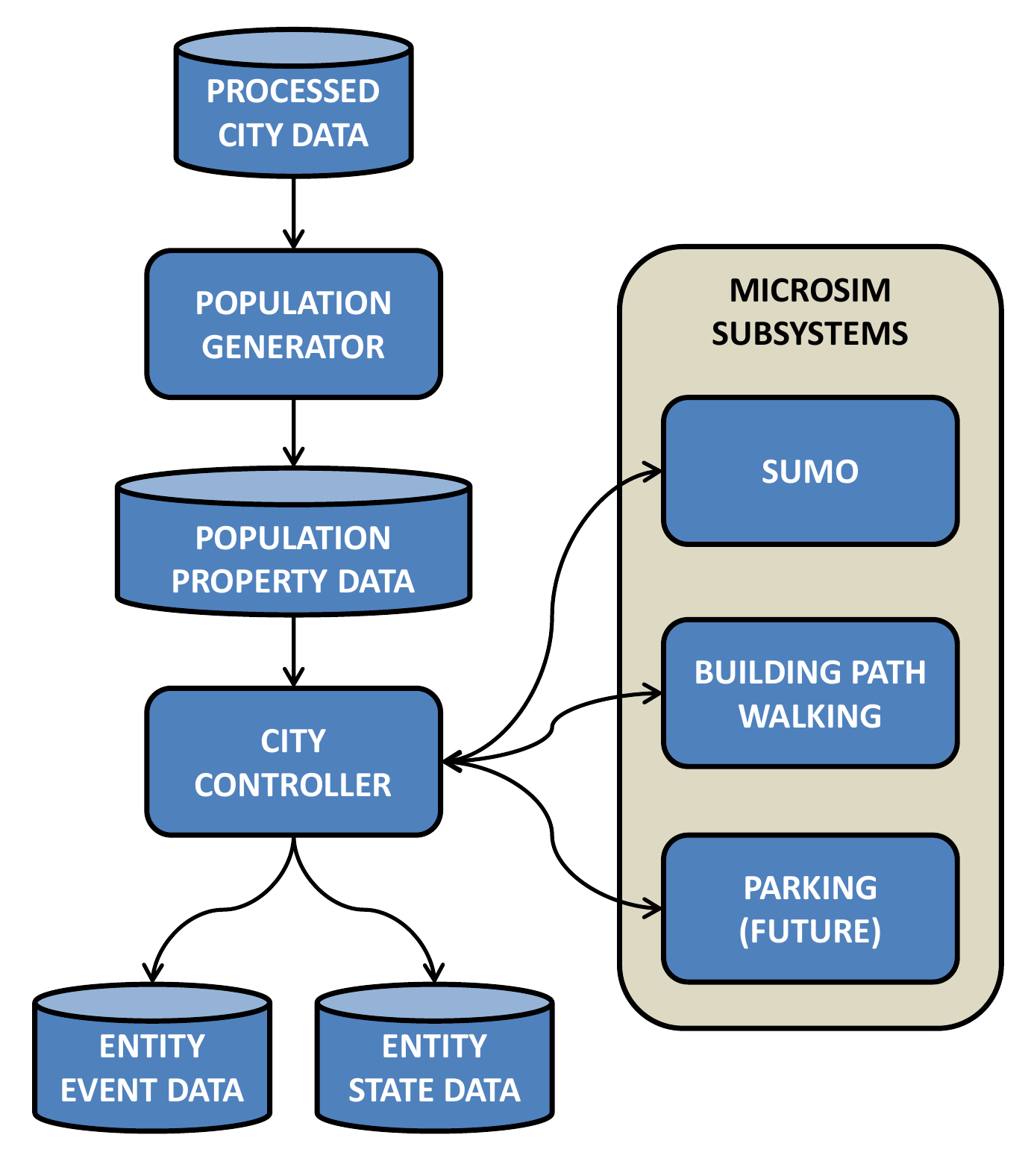}
\caption{Internal system components for the City Simulator. The city population is generated by random number of people to each house, and are also assigned places to work, shop and recreate.  This co-habitation and common meeting places stimulates a pattern of life where people can be associated together.  The simulator employs a number of subsystems to perform traffic simulation (SUMO), walking to and from buildings and routing and loading/unloading of buses, external to SUMO.  The entity state of all vehicles and people is stored at each timestep, and significant events such as entering a building are recorded for fast processing of metadata.}
\label{fig:city_sim_system}
\end{figure}

This paper proposes a framework for the generation of simulated city-wide image/video data; incorporating the three-dimensional terrain, buildings and road data, modeling the movement of people and vehicles, pattern of life information for individuals and locations, and a high resolution visualization tool that can be used to generate long duration high resolution video data across the city area. The proposed framework is outlined in Figure~\ref{fig:full_system_framework}. It adopts a similar approach to the wide area image generation as the US Army PerSEval project~\cite{deaver2010perseval} but it takes a person-centric approach, with an individual's route planning, interactions and individual intent being modeled explicitly to provide a more natural representation of the pattern of life. The paper is organized as follows: Section~II describes how objects traverse the city simulation, switching seamlessly between different simulation engines;  Section~III describes the image generation and configurations of wide field of view cameras; Section~IV covers associated querying and playback tools for city metadata and the large video dataset; Section~V concludes the paper with a discussion on future expansion of the system.

{\section{Related Work}\label{fig:chapter1a}}

\noindent Most work in the area of traffic modeling and the simulation of pattern of life is aimed at producing an environment where entities or agents can move and interact in a realistic manner and that demonstrate emergent properties that can be studied and related to behaviors in the real world. In particular, with the rapid growth of computing power, there has been a significant improvement in the fidelity of simulations of traffic~\cite{Thomin2013urbantraffic,Brugmann2014realworldtraffic,Thonhofer2018macrotrafficnetworkdesign,LopezNeri2010microscopictraffic,Zehe2015cloudbigdata} and pedestrians~\cite{Vogt2012pedestrian}. In addition, the ability to model vehicles in a complex traffic network allows for the improvement of other vehicle simulations, such as driving simulations with realistic vehicle controls and good graphical rendering of the three-dimensional scene for the human operator or driver~\cite{Sun2015drivingsims}. The aim of the current paper is to use some of these approaches to populate a large urban or city environment that contains three-dimensional terrain features, real background imagery of the ground, three dimensional buildings, a realistic road network and traffic simulation, and individual people (pedestrians) who are associated with a series of tasks as part of a simulated pattern of life. Although there has been a lot of previous work on developing algorithms to track the movement of vehicles and individuals through such a complex environment, progress in this area has tended to rely on the availability of real image data and a small number of large scale trials involving real sensors. the current work provides an alternative means to generate realistic looking surveillance and reconnaissance data, and the ground truth to be used to evaluate the performance of the algorithms.

The data used in the present work (three-dimensional terrain, background features, road layout and building information) are all derived from open source data sets, whilst the traffic and pedestrian modelling uses a standard and a widely-used open-source simulator. There are a number of established open-source vehicle simulation tools available which operate at different levels of detail and abstraction, divided into the following classifications~\cite{burghout2005mesoscopic}: Macroscopic simulation - examples of which tend to consider roads load solving as a form of conservation of a quantity entering and leaving the road ``pipe", these simulations are applied to strategic planning applications~\cite{Thonhofer2018macrotrafficnetworkdesign}; Mesoscopic simulation - a hybrid approach where microsimulation may be used within the road ``pipe" but detailed behaviors such as waiting at junctions to turn etc, may not be present; Microscopic simulation - involves the simulation of individual vehicles traversing the road network and interacting with other vehicles, to variable degrees of fidelity~\cite{LopezNeri2010microscopictraffic}.  An example macroscopic simulation is MASTER~\cite{helbing2001master} (a simulation based on gas-kinetic equations), MATSim~\cite{horni2016multi} is an example of a mesoscopic simulation, and available microscopic simulations are MovSIM~\cite{treiber2013traffic} and SUMO~\cite{behrisch2011sumo,krajzewicz2002sumo}.  The goal of video generation can only be achieved with a microscopic model, both are peer reviewed models however SUMO was selected over MovSIM as it has a larger and more established acceptance in the transport simulation community, and its toolset has a longer history of use. However, the approach adopted here could easily be modified to incorporate MovSIM or another traffic simulator as an expansion of its capability.

{\section{Pattern of Life Simulation}\label{fig:chapter2}}

In order to provide a plausible set of moving targets and background clutter the system uses a simulation of a city with the aim of creating a pattern of life movement between buildings~\cite{rekimoto2007lifetag}.  This is accomplished through a system (Figure~\ref{fig:city_sim_system}) that uses a MATLAB controller to interact with a traffic simulation tool (SUMO).  The controller contains a number of state machines that manage the methods that people use to traverse the city, the controller outsources the microsimulation of people and vehicles where appropriate, merging together the result and transitioning to the next stage of the travel plan.  It also ensures significant events such as leaving/entering a building and boarding a vehicle are consistently recorded and timestamped for analysis.  The city structure (buildings and road networks) is prepared from the open data sources such as the Open Street Map project~\cite{osm_website}, added to this data are autogenerated building footpaths, roadside waiting and pickup points, and automatic association of these roadside positions with buildings far from the road network.

The city inhabitants are each allocated at random one house, workplace, preferred shop, and preferred recreation building.  The housing for a proportion of the population (25\%) is reallocated to form co-habitation relationships with the remaining population.  Vehicle are assigned to a proportion of the population (50\%), people who have a personal vehicle will drive directly to the destination as soon as a road becomes accessible to the person. People without vehicles will either walk to the destination if within an acceptable range, else will walk to a bus stop to commute (if no advantage can be gained from a bus journey the person will walk direct to the destination).   

Task assignments, such as ``Go to work'' or ``Go shopping'' are randomly allocated to each inhabitant in adjustable proportions. These tasks are implemented through trips (journeys) where the endpoints are extracted from the person's metadata, for example, travel between the locations of the person's home and workplace. The trip is allocated a start time, distributed around a mean start time for the task. The traffic flows created by these tasks could potentially be matched with corresponding traffic data, to determine more representative start times and task assignment proportions. In addition to general purpose workers (which visit and occupy workplace buildings during work hours), a number of bus drivers are assigned to the population as specialist occupations that stores and manages a bus route

{\subsection{City Generation}\label{sec:citygen}}

\begin{figure}
\centering
\includegraphics[width=\columnwidth]{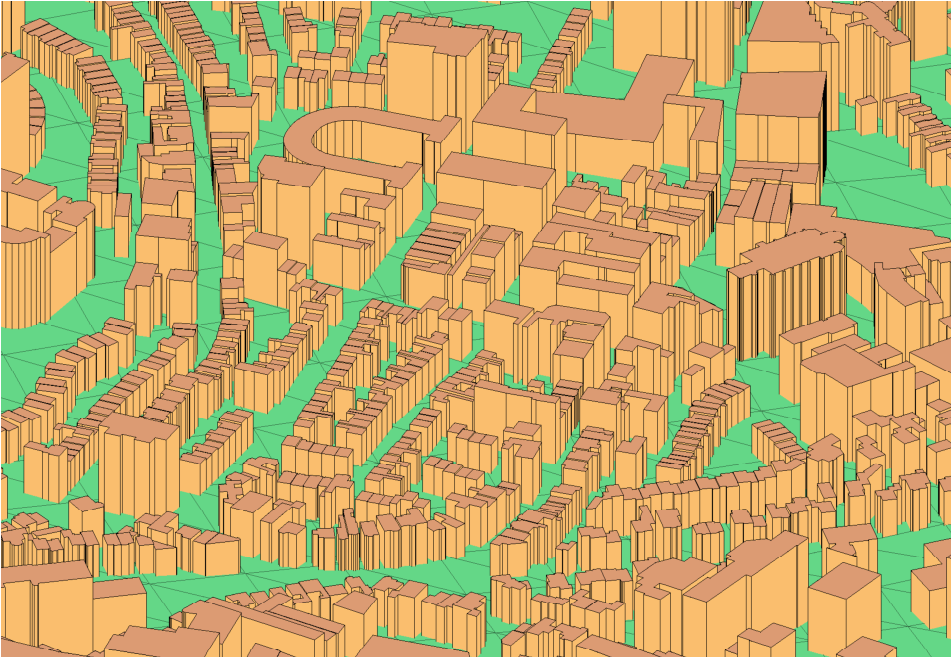}
\caption{Building geometry.  In the absence of known building heights, the buildings are assigned a random height within a range constrained by the building type, and scaled by the floor area of the building.}
\label{fig:3d_buildings}
\end{figure}

\noindent Three dimensional models of the city buildings are extruded from base polygons extracted from Open Street Map data (Figure~\ref{fig:3d_buildings}), the unknown building heights can either be autogenerated from the building floor area (polygon area) scaled between minimum and maximum heights, defined by building type in MATLAB, or alternatively through a similar technique, using the OSM2XP~\cite{OSM2XP} tool to generate buildings in the X-Plane flight simulator, which has been used as the main visualization tool for the current system. These building heights are then synchronized with City Simulator by parsing the text file generated by the X-Plane ``DSF2Text" scenery conversion tool.  The buildings and roads are positioned on a 3D terrain mesh, and as with the buildings, this terrain can either be generated randomly, processed from an external heightmap or extracted from the X-Plane flight simulator and imported.    
The system uses the Simulation of Urban MObility (SUMO)~\cite{behrisch2011sumo} to simulate the vehicular and pedestrian traffic (the usage is elaborated on in sections~\ref{sec:sumo} and~\ref{sec:microsim}) with the road network being converted from Open Street Map (OSM) source data to the SUMO road network specification using the supplied SUMO ``NETCONVERT" conversion tool.  The conversion process generates internal road and junction lane geometry, either from the number of lanes specified in the OSM data, or estimated from the specified road type.

\begin{figure}
\centering
\includegraphics[width=0.8\columnwidth]{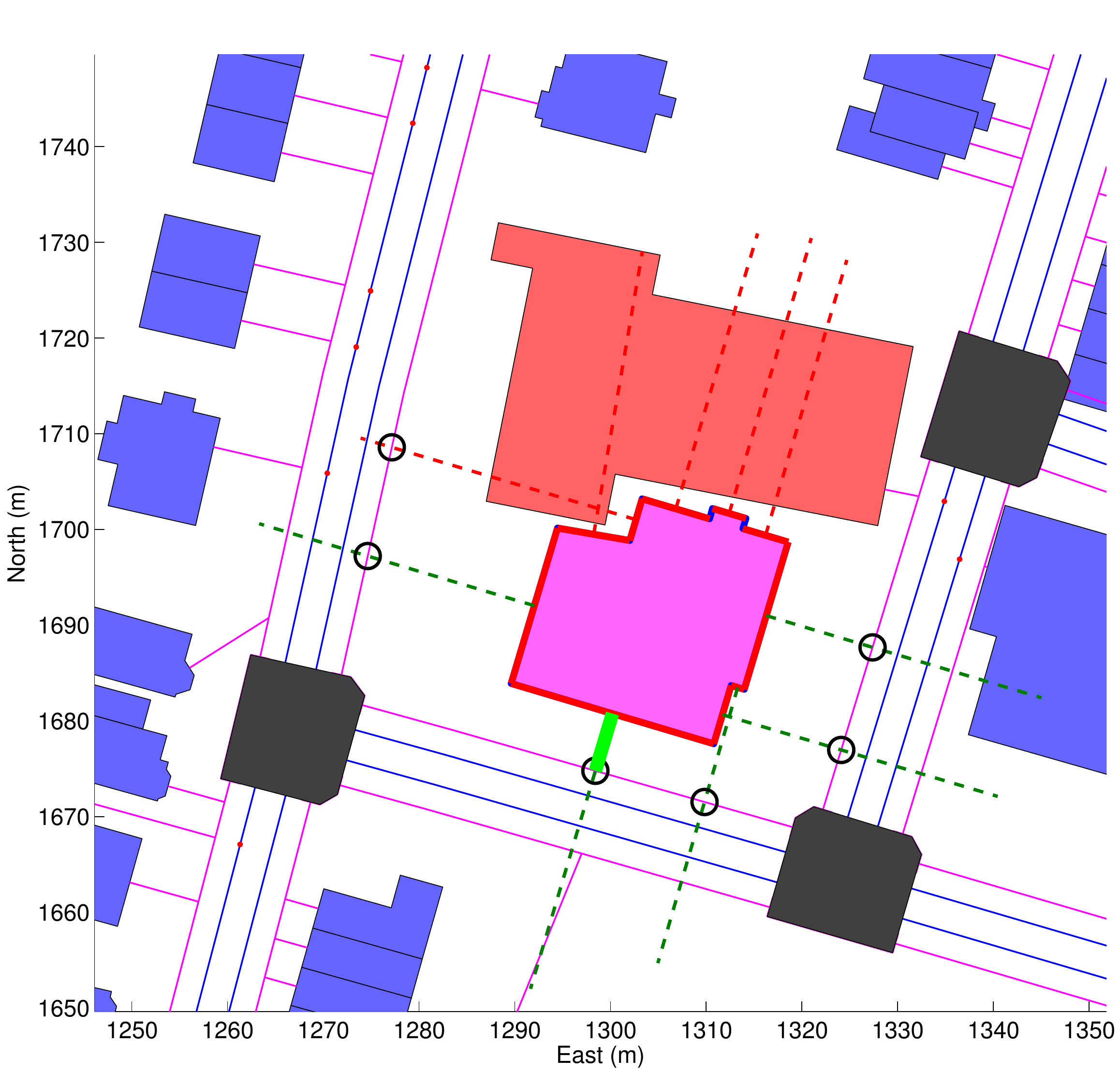}
\caption{The process of autogenerating building footpaths extends probes (dashed lines) perpendicular to every wall, to intersect with the path network (purple lines).  Probes are only created from walls long enough to support a doorway.  If a probe intersects with another building wall or exceeds 30 metres in length without intersecting, it is discarded (red dashed line).  Intersections with junctions are considered failures to prevent vehicles stopping mid-junction.  If the probe successfully intersects with the path network (thin purple line), the probe is stored (black circles), and the probe with the shortest length is used as the building footpath (thick green line).  The chosen intersection point is used as the building's gateway, and the probe is further extended to find the intersection with the road network (thin blue line) which becomes the vehicle stopping point for this building.}
\label{fig:footpath_generation}
\end{figure}
\begin{figure}
\centering
\includegraphics[width=0.8\columnwidth]{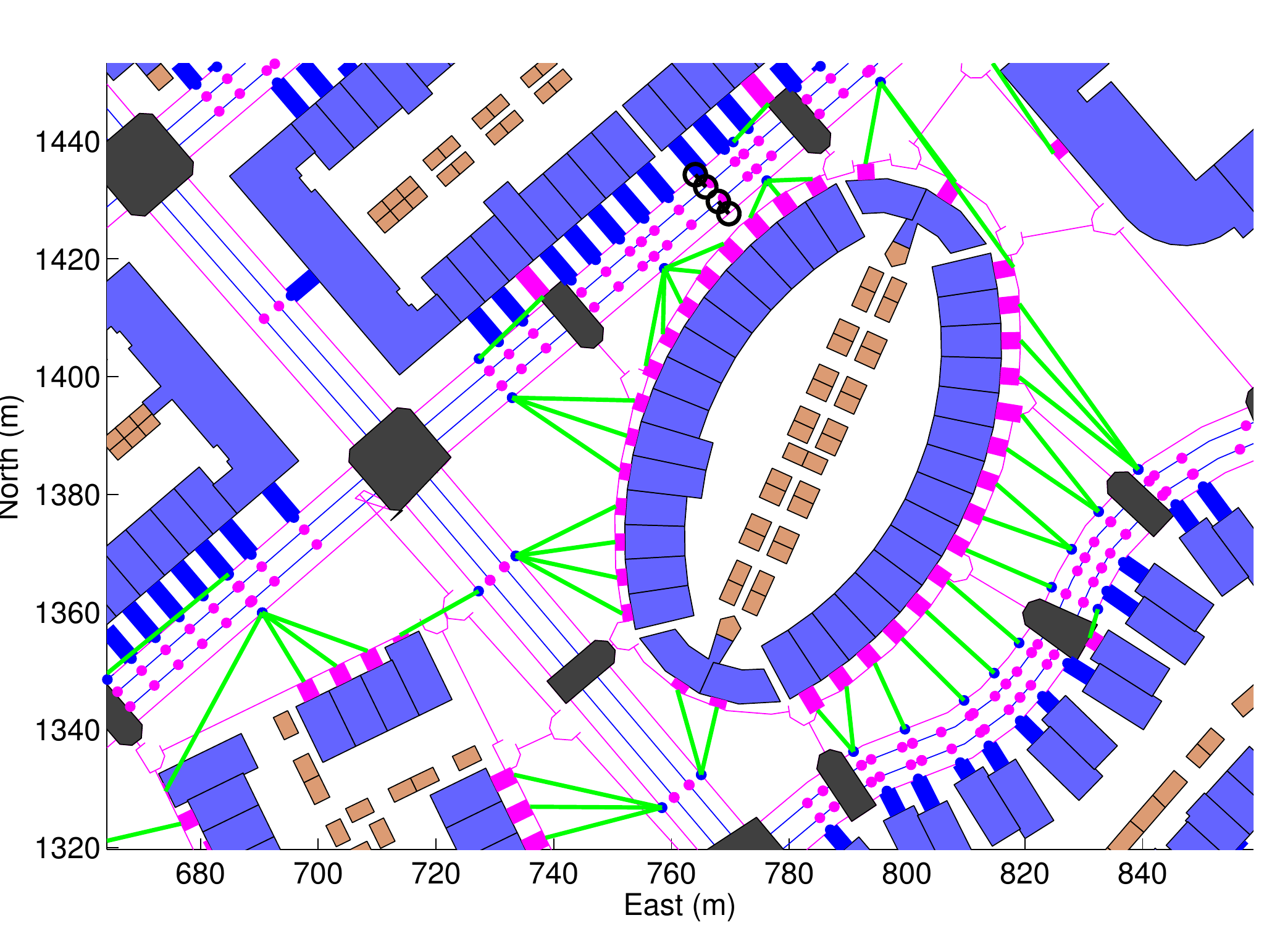}
\caption{Map showing building footpaths (thick blue and purple lines) and vehicle stopping points (purple dots) and pedestrian waiting points (blue dots).  For buildings whose footpath has no direct access to the road (thick purple lines), additional vehicle-person transfer points are generated and associated with building gateways.  New stopping points are generated from the midpoints of road line segments, and linked to the nearest gateway by Euclidean distance (green lines).  The linkages shown are symbolic and the SUMO system will walk the person using the path network (thin purple lines)}
\label{fig:footpath_linkages}
\end{figure}

\begin{figure}
\centering
\includegraphics[width=0.8\columnwidth]{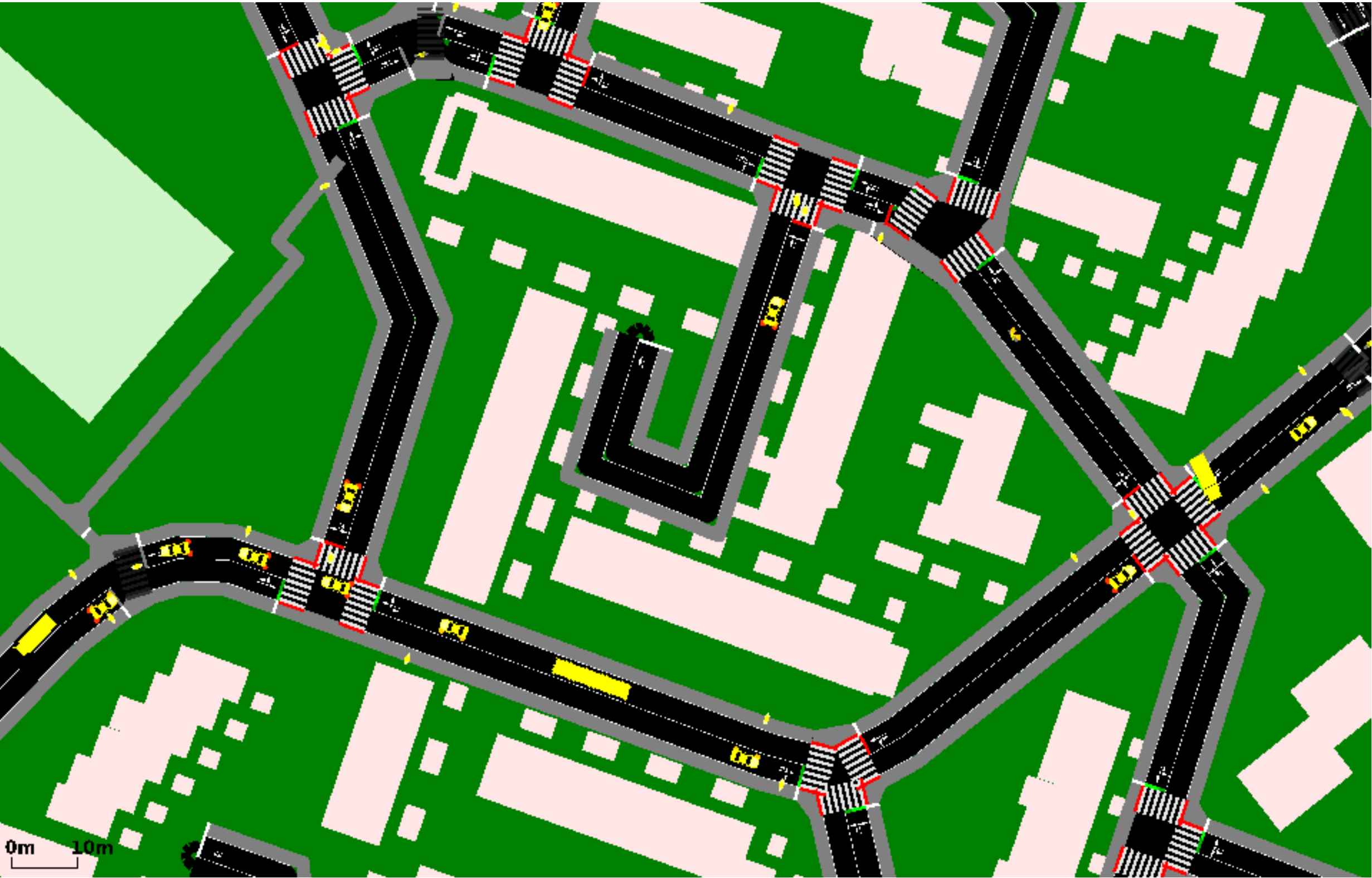}
\caption{SUMOGUI (part of the official SUMO package) showing the simulation of mixed traffic on a road network with footpaths and sidewalks.  The buildings shown here are displayed polygons and have no physical interaction with the simulator, building footpath walking (from roads to building doorways) is external to the SUMO simulation. }
\end{figure}
\begin{figure}
\centering
\includegraphics[width=0.8\columnwidth]{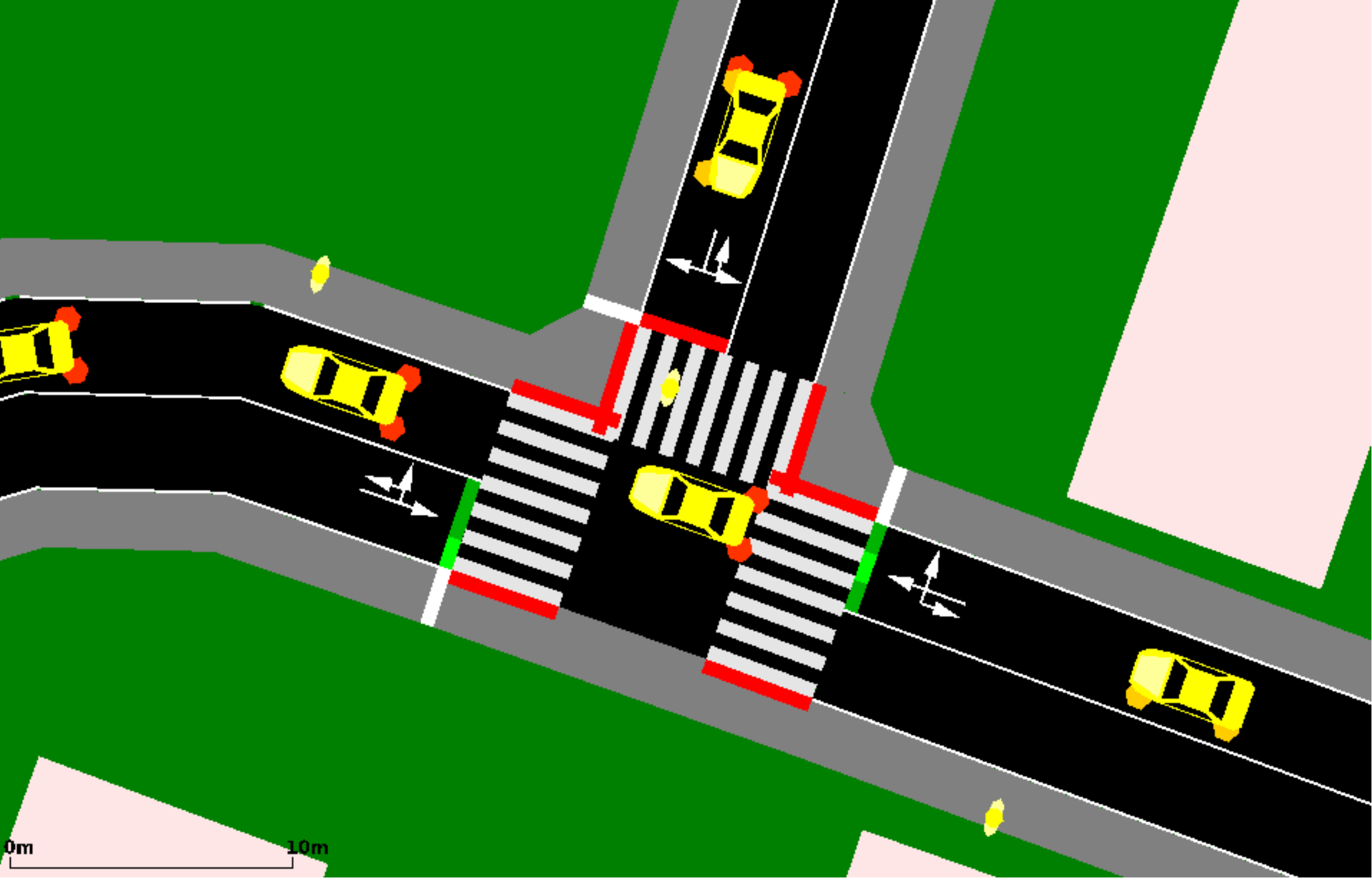}
\caption{Magnified extract showing individual pedestrians walking and crossing, with traffic lights and vehicle indicators. Unlike macroscopic and mesoscopic simulators, the microscopic SUMO will simulate smooth turns through these junctions.}
\label{fig:SUMO}
\end{figure}

The SUMO model provides sidewalks adjacent to the roads, but it does not automatically provide footpaths from the sidewalks to the buildings. A \textit{building footpath} tool is used to permit access to each building from the \textit{path network} thereby allowing the continual simulation of the person from door-to-door.  Since the footpath geometry is not available from open sources, it has been autogenerated (Figure~\ref{fig:footpath_generation}).  The autogeneration procedure starts by checking the length of a wall segment to determine if it is wide enough to support a \textit{doorway}.  If there is sufficient width, a candidate doorway is placed halfway along the wall.  Next a test ray is projected from the doorway outwards perpendicular to the wall, if this ray intersects with (reaches) the \textit{path network} before intersecting with another wall (blocked) it is considered as a potential footpath.  After all the walls have been checked the shortest length footpath is used for accessing the building.  

The selected footpath's intersection point with the larger path network is used as a \textit{gateway}, and if the footpath ray intersected further into a neighboring road lane, this becomes the vehicle's stop position, and the gateway becomes the person's wait position, for a particular \textit{vehicle-person transfer point}.  

If no road intersection occurred during the footpath generation phase, then the building is considered to be indirectly coupled to the road network via a pedestrianized area, and therefore requires a remote location for the \textit{vehicle-person transfer point}.  People using this building will walk from the \textit{gateway} to this point via the larger path network prior to accessing any vehicle (and vice versa).  To ensure transfer points are available, additional sets are generated at the midpoints of all road segments (that have a sidewalk) in the entire network.  The vehicle-person transfer point nearest (by Euclidean distance) the building's gateway point is selected as shown in Figure~\ref{fig:footpath_linkages} by the green linkage lines.

\begin{figure}
\centering
\includegraphics[width=\columnwidth]{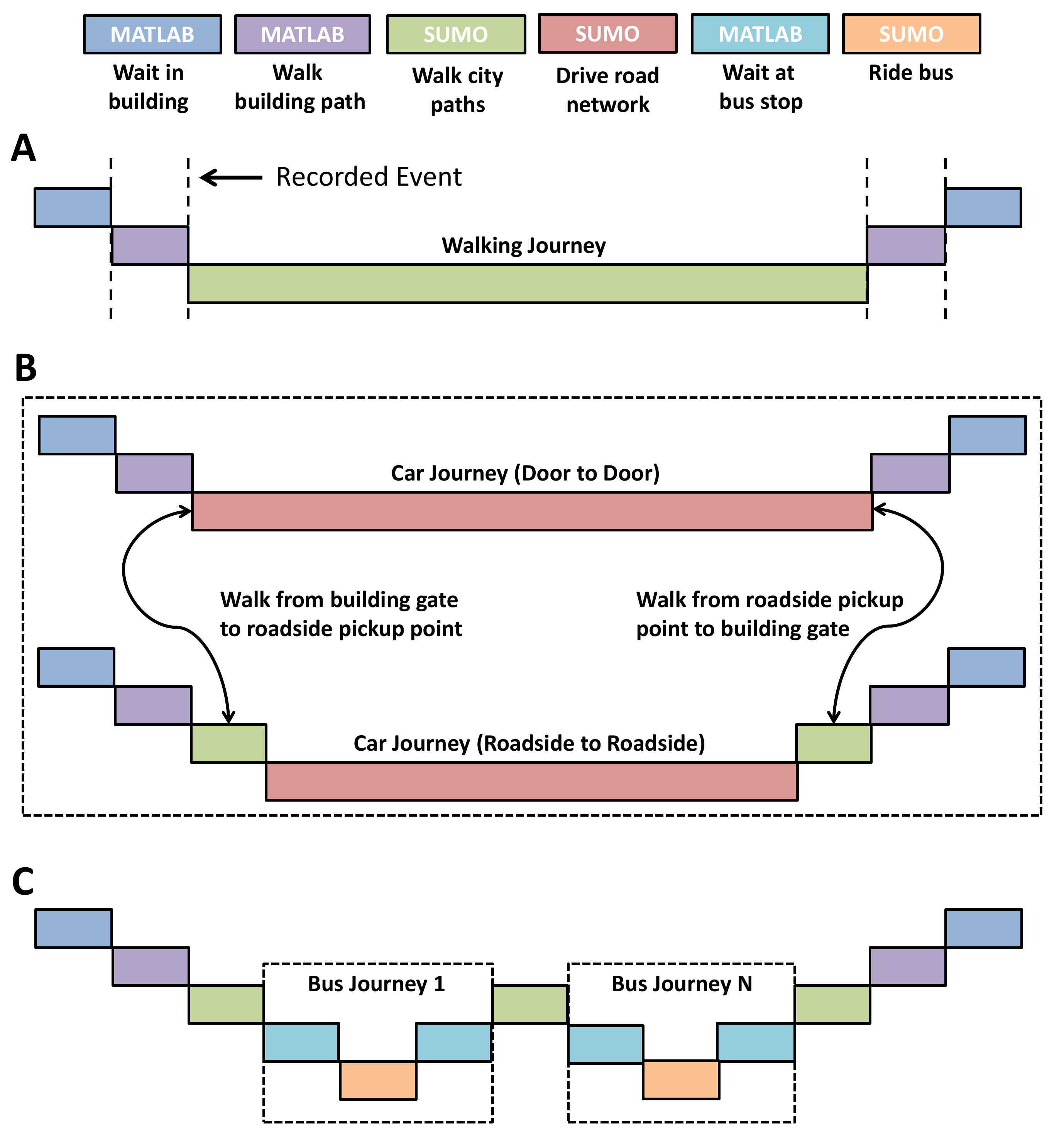}
\caption{\textbf{(A)} Travel sequence for walking directly from building to building. \textbf{(B)} Travel sequence for vehicle journeys, the second option includes an intermediate walk to or from a building that is far from the roadside. \textbf{(C)} Travel sequence for a multistage bus journey, with a walk between each bus stop.}
\label{fig:travel_sequences}
\end{figure}

\begin{figure}
\centering
\includegraphics[width=0.8\columnwidth]{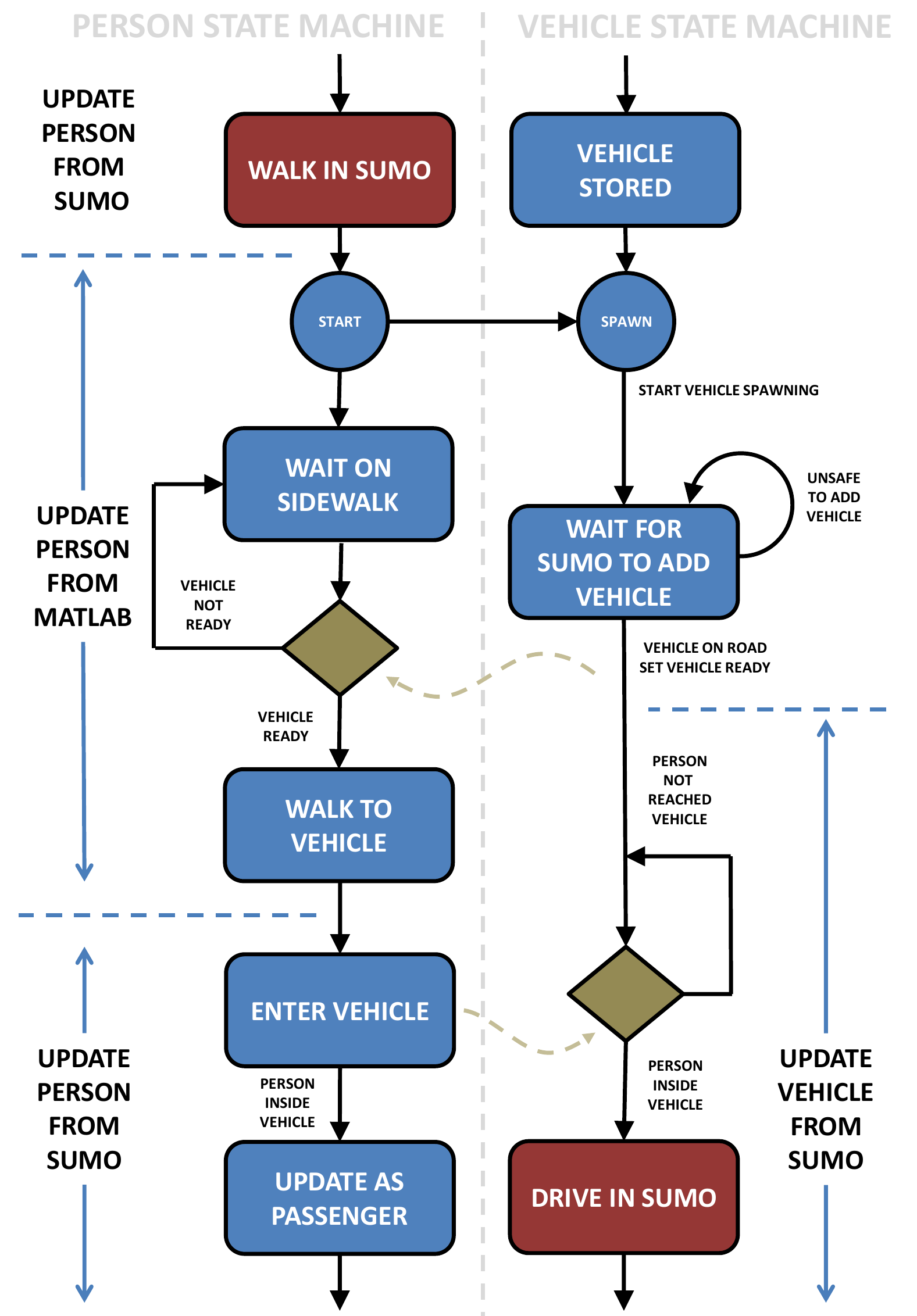}
\caption{Example of a state machine interaction.  This example shows a person walking to the roadside, spawning a vehicle, waiting for the vehicle to be ready, accessing the vehicle and driving the vehicle.  The diagram shows where SUMO (red boxes) and systems outside SUMO (blue boxes) are involved.  The diamond boxes are wait states released by an event (curved dashed arrow) in another state machine.  The regions defined between the horizontal dashed blue lines indicate where the positional update is currently sourced for both the person and the vehicle.}
\label{fig:wait_for_vehicle}
\end{figure}

\begin{figure}
\centering
\includegraphics[width=0.8\columnwidth]{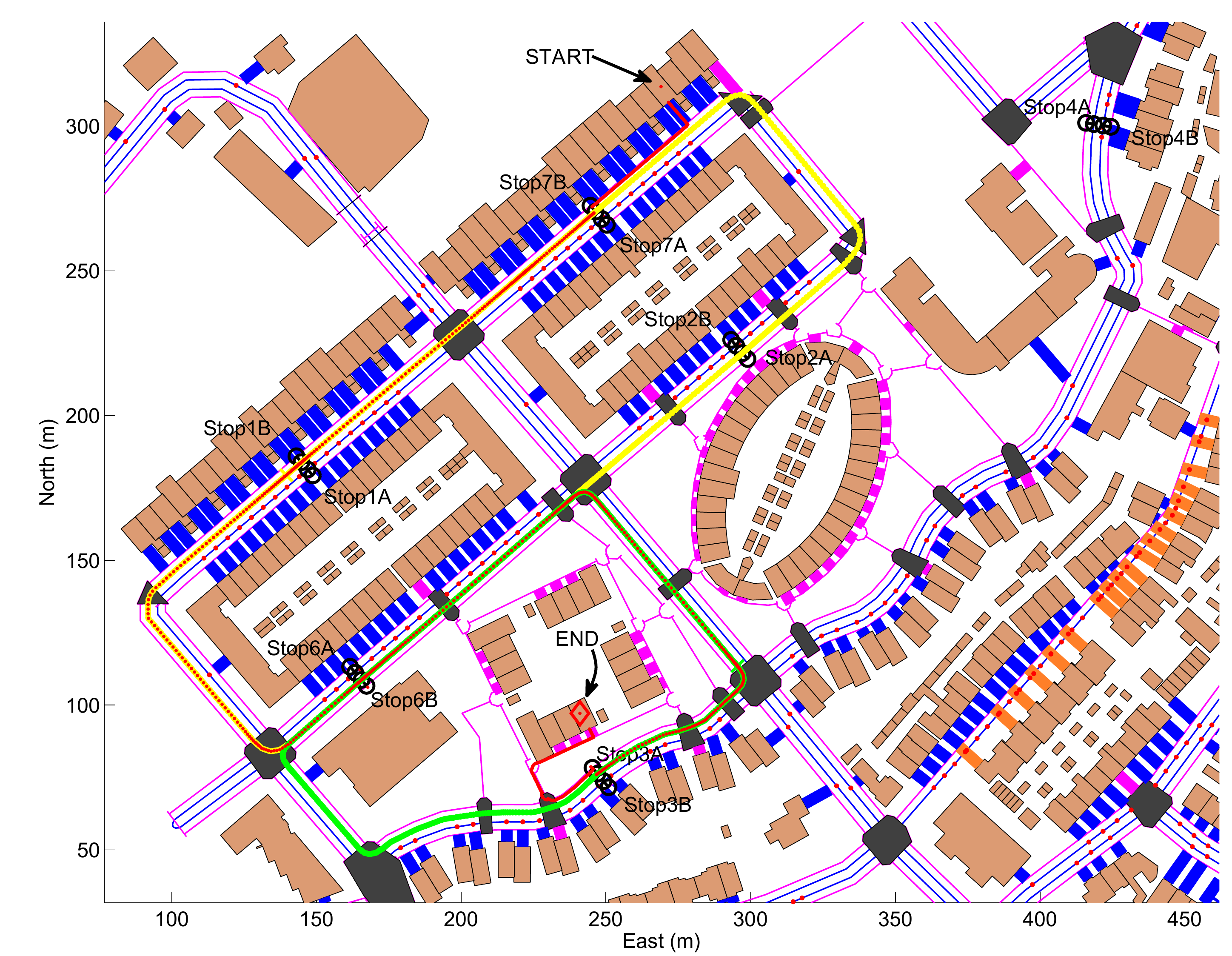}
\caption{Map showing example trajectories of one person using two buses (the bus routes are small loops for demonstration purposes) with the goal of minimizing walking distance.  The person (red line) starts in a building, walks the building footpath, walks to the bus stop 7B and waits for the yellow bus.  The person boards the stopped bus, and then the bus continues to stop 6B, where the person waits for a second bus to stop 3A (the person will walk to a second bus stop if the second bus route used different stops).  The person then ceases to use bus routes, as no advantage is gained from using stop 3A, and so the person is then routed from the bus stop to the gateway of the building, finally walking the building footpath into the building.  Of particular note is the separation of the red dots when the person is transported by the bus showing the acceleration and deceleration at junctions and bus stops.}
\label{fig:example_traj}
\end{figure}

{\subsection{SUMO Traffic Simulator}\label{sec:sumo}}

\noindent    In this work, the open source traffic simulator known as `Simulation of Urban MObility'  or `SUMO' has been used and adapted to reflect the needs of the WAAS simulation~\cite{behrisch2011sumo,krajzewicz2002sumo}. SUMO has been available since 2001 and was developed by the Institute of Transportation Systems at the Deutsches Zentrum f\"{u}r Luft und Raumfahrt (DLR) for evaluating modifications to road infrastructure and transport policy, examples are: optimizing traffic light timings~\cite{krajzewicz2005simulation}, forecasting traffic density~\cite{liang2014real,habtie2016neural} and evaluating wireless in vehicle systems known as ``Vehicle-to-X" (V2X) infrastructure~\cite{uppoor2014generation} .   

SUMO models the motion of individual vehicles and pedestrians, calculating positions of these entities at each simulated timestep, it also models the interactions between vehicles and pedestrians at road crossings (Figure~\ref{fig:SUMO}).  Internally SUMO supports a wide variety of vehicle following models, ranging from the Krau{\ss} Driver Model~\cite{krauss1998microscopic} (as the default model) to the well known Intelligent Driver Model (IDM)~\cite{treiber2000congested} and variants there of.

Control and querying of the SUMO system is possible through TraCI (Traffic Control Interface)~\cite{wegener2008traci}, a UDP packet based interface that is programming language independent. TraCI does provide an interface to Matlab via the~\textit{traci4matlab} API~\cite{acosta2015traci4matlab}). Unfortunately, MATLAB has relatively poor performance when typecasting between datatypes and therefore a bespoke MEX file was used to replace the~\textit{traci4matlab} function that decodes incoming UDP bytes into MATLAB structures. 

The system presented here implements a person-centric simulation of a city, however at the time of implementation (SUMO 0.25) the TraCI specification did not have a mechanism to add or route new pedestrians, therefore the TraCI server in the SUMO C++ code was modified to replicate the vehicle interface, but for pedestrians.  This had the key benefit of the SUMO simulation treating pedestrians as people, and so interact properly at traffic crossings with vehicles.  Current iterations of SUMO now have a similar, officially supported, TraCI pedestrian interface (unrelated to this project). SUMO also includes a model for public transport (buses), however TraCI does not yet support control over the loading and unloading of passengers, therefore this functionality has been replicated at a higher level as part of the MATLAB controller's state machine.  The buses are manually controled vehicles, with routing and stopping at bus stops controled by standard TraCI~\textit{Vehicle State Change} functions.  This has the additional advantage that a more complex bus controller could be added without changing the core of SUMO, and allow interaction with 3rd party tools and other external subsystems of variable complexity.  For example, a controller could be added that attempts to maintain a fixed bus timetable, that also interact more realistically with a person task planner that is now timetable aware and can account for waiting times.

Although the SUMO software provides a dynamic model for the traffic flow and interaction between vehicles, it is still ultimately limited by the data used to define the parameters relating to the number of vehicles using the roads and the typical journeys being made. The model developed here is not intended to be used for traffic modelling per se, but it is often useful to have a mechanism to introduce some realism into the system. To do this, a traffic junction model has been developed, based on the based on the SCOOT traffic sensors used extensively in the UK and elsewhere internationally \cite{Bretherton1990237,hounsell1990scoot}.  The model provides a means to increase or reduce traffic flow locally (a stochastic source-sink process) to change the traffic rate to reflect the time of day, and when it is applied to junctions on the periphery of a large area, it can help to increase the realism within the simulation without causing too many unwanted effects. For example, the use of this in WAAS modelling could lead to the creation/removal of vehicles from the road network as being `anomalous' behaviour and this could negatively affect any pattern of life extracted from the WAAS image data.  
\begin{figure}
\centering
\includegraphics[width=\columnwidth]{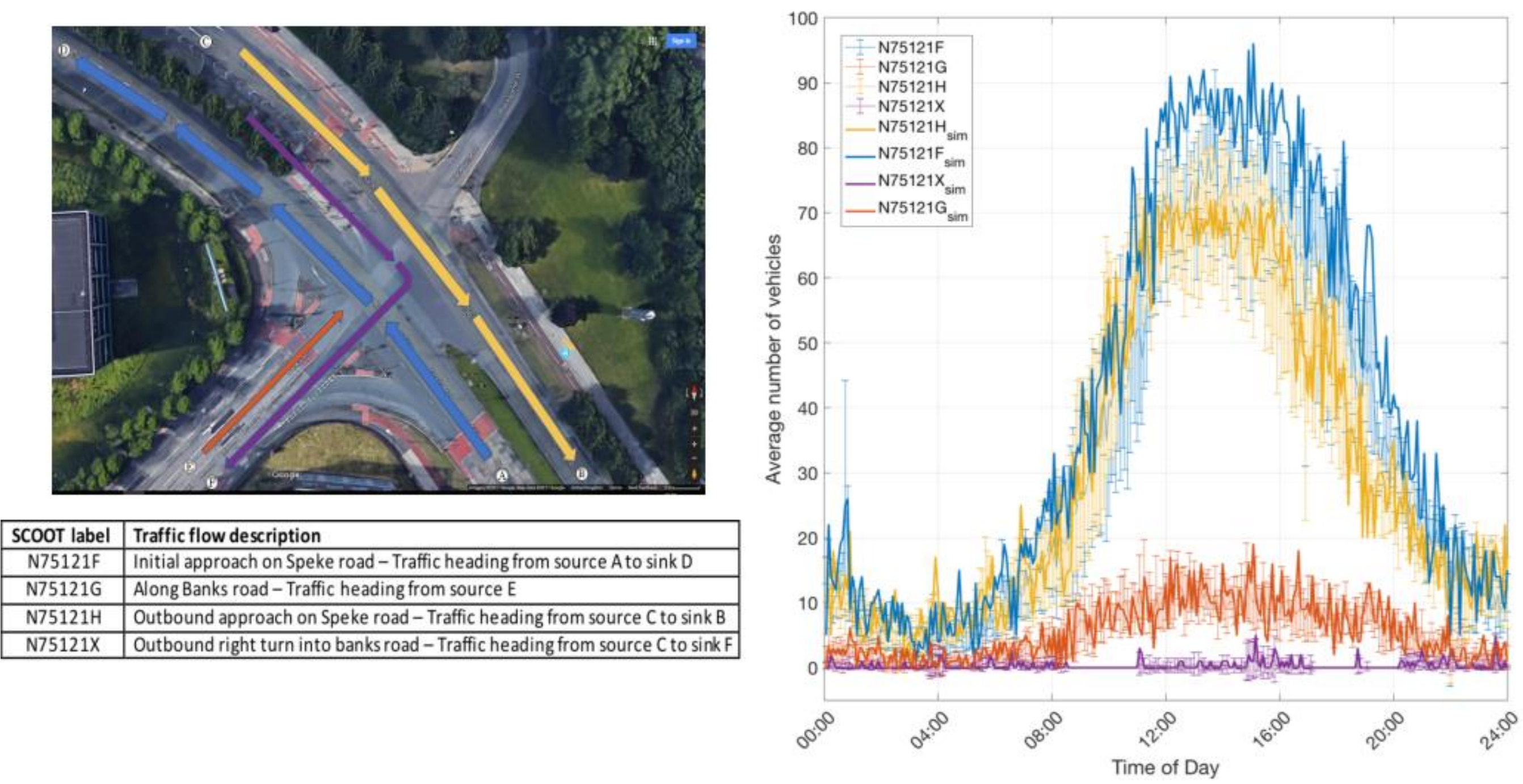}
\caption{An example SCOOT monitored junction in Liverpool (at latitude 53.350530 and longitude -2.882987) with the different junction elements listed (left), and a comparison of the simulated traffic flow and the real averaged traffic flows for weekend traffic as a function of the time of day (right).}
\label{fig:scoot}
\end{figure}

In practice, the number of vehicles and people moving around the scene is limited by the computational resources available to the simulation. The methods outlined here are intended to generate representative video imagery for large areas of a city, up to 40 km$^2$ for the largest WAAS sensors, and the current system has been demonstrated for simulations of up to 100,000 individual vehicles (with four different types of vehicle) and 50,000 pedestrians.

{\subsection{Microsimulation and Travel Sequences}\label{sec:microsim}}

\noindent The simulation controller generates and executes a number of state machines that implement the sequence of events required to traverse the city.  In each state a microsimulation is performed, examples of such states are ``DRIVE\_TO\_POINT" or ``ENTER\_VEHICLE".  Whilst in each state, every simulation cycle updates the entity position and status using the behavior defined by the state and interacts with other entities as required (e.g. update passengers).  The state is transitioned when the completion criteria is satisfied, for example, reaching the transfer point completes a ``DRIVE\_TO\_POINT" state and proceeds to ``EXIT\_VEHICLE", and then perhaps a ``WALK\_TO\_GATEWAY" may begin.
	
Simplified examples of chained states are shown in figure~\ref{fig:travel_sequences}, with detailed implied behaviors such as entering and exiting vehicles removed for clarity.  Each state transition is logged with a timestamp, a location and a set of metadata specific to the event.  
Figure~\ref{fig:travel_sequences}A illustrates walking directly from the starting building to the destination building, this can occur if either: the destination is close to the start point; the person does not own a vehicle and the bus network does not provide any advantage.  In MATLAB, the person waits at the start building until time to begin the journey, walking the building pathway from the building doorway to the building gateway (city path network), the person is then transferred to the SUMO simulation to walk the city path network.  The sequence is reversed when the person reaches the building gateway, transferring back to MATLAB the person walks to the destination doorway and enters the building waiting for the next journey.  
Figure~\ref{fig:travel_sequences}B shows two variations of traveling in a personal vehicle, and occurs when a person owns a vehicle and the destination is far.  Figure~\ref{fig:travel_sequences}B (top) involves transport of people direct from the gateways of the start and destination buildings, whilst figure~\ref{fig:travel_sequences}B (bottom) occurs if one or more of the building gateways do not have direct access to the road network and requires a walk stage to be inserted.  
Figure~\ref{fig:travel_sequences}C demonstrates multiple uses of bus routes to reach the destination, an example is shown in Figure~\ref{fig:example_traj}.  The MATLAB controller manages waiting at bus stops, boarding and disembarkation.  If the person can gain an advantage using a further bus, the person walks to the next bus stop (in SUMO) to repeat the cycle until the destination building is reached - else the person walks to the gateway of the destination building and completes the journey.
  
The microsimulation occurs in the Universal Transverse Mercator (UTM)~\cite{chang2006introduction} coordinate system, a flat Cartesian local projection of the Earth, parametrized by a Proj4~\cite{proj4} string generated by the SUMO NETCONVERT during the processing of the Open Street Map data (which is defined in geographic latitude and longitude).  The Proj4 map projection library is also used to convert City Simulation positions back to geographic latitude, longtiude and altitude, as the image generator operates in that global coordinate system.
	
\subsubsection{Bus route selection}		
		
\noindent To determine a robust (but not necessarily optimal) set of bus stops to use, equation~\ref{eqn:bus_stop_metric} is solved:
\begin{equation}
	\argminl_{m,n,r}{\left(\norm{\overrightarrow{start~to~stop_m^r}} + \norm{\overrightarrow{stop_n^r~to~dest}}\right)}\\
	\label{eqn:bus_stop_metric}
\end{equation}
\noindent where $m$ is the boarding bus stop index and $n$ is the index of the disembarkation bus stop and $r$ is the route index where the maximum values of $m$ and $n$ can change with $r$.  Using this nomenclature $\overrightarrow{start~to~stop_m^r}$ is the vector from the (current) start position to boarding bus stop $m$ and is an estimate of the effort to walk to that bus stop.  Likewise, $\overrightarrow{stop_n^r~to~dest}$ is the vector from the disembarkation bus stop $n$ to the destination building and is an estimate of the effort to walk from that bus stop.  Care must be taken to constrain the set of $n$ stops to only those on the route(s) $r$ that leave from bus stop $m$.  

To solve for multiple combinations of $m$, $n$ and $r$ (which represents a journey using multiple buses) is more complex, and involves a recursive branching search from each disembarkation bus stop by setting the current $stop_n^r$ position to be the new $start$ position, for repeated solving of Equation~\ref{eqn:bus_stop_metric} to obtain a global minimum.

To determine the best bus travel direction to use, the number of stops between the chosen bus stop pair is then minimized.  If the goal is to fully optimize bus usage, more accurate decisions could be made by using the actual length of routes to the bus stops, and the bus route itself, however in the current surveillance use-case (of creating entities that can be tracked), generating a subset of sensible bus journeys is sufficient and bus route optimization is beyond the scope of this current work.

SUMO performs the micro-simulation for walking the path network and the driving along the road network, people and vehicles are added to the SUMO simulation only when required, and are removed when the walking or driving segment of the journey is completed.  This complete removal of entities from SUMO, whilst maintaining the overall state in the MATLAB controller, minimizes the processing requirements on SUMO for otherwise maintaining inactive (waiting at bus stop, in building, in vehicle) entities.  

Walking from the building to the SUMO network was performed in MATLAB with a path following function that updates the position of a fixed velocity particle traveling along a line.  This avoids the alternative of modifying the SUMO network, which would require additional paths and the splitting of existing paths for the placement of new junctions, which would greatly increase the complexity of the simulation for what is essentially a small well defined process.  The hybrid approach of blending MATLAB footpath walking with SUMO path network walking is also a proof of concept for the integration of more complex behaviors and motions outside of SUMO.

Figure~\ref{fig:wait_for_vehicle} shows an interaction between person and vehicle state machines, via MATLAB (blue boxes) and SUMO (red boxes).  For each timestep, all people and vehicle objects in the system are updated, this can vary from a costly action such as adding a vehicle to SUMO (usually only when entering a state) to a simple check if a wait state should be released.  The flow diagram shows an example of a person completing a ``WALK\_IN\_SUMO" state, and entering a ``WAIT\_FOR\_VEHICLE" state.  On entering this wait state a vehicle is spawned, and the vehicle state machine begins to wait for the car to be inserted into the traffic in SUMO as a parked vehicle (current SUMO behavior is to only add a vehicle if safe to do so).  The person state machine exits the wait state when the vehicle signals it is ready (by raising the flag in the vehicle object the person is waiting on) and then the vehicle enters a wait state.  The person then walks from the sidewalk into the road in order to enter the vehicle, when completed the person is added to the vehicle's passenger list which will update the person's state with the vehicle state until the person leaves the vehicle.  Finally, the wait state on the vehicle is released and the waiting (parked) SUMO vehicle is commanded to start driving to its destination (transfer point).

In cases where there is multiple occupancy in a building, there is also the option of the occupants of the building to share a journey and drop off the passengers at multiple stops, making more efficient use of vehicles. This is implemented through the same mechanisms as the Bus routes, but with a lower passenger capacity of 5 people, and the bus stops being the building drop-off points. The list of stops along the route are then dynamically set by the list of whatever passengers in the car.

\begin{figure}
\centering
\includegraphics[width=\columnwidth]{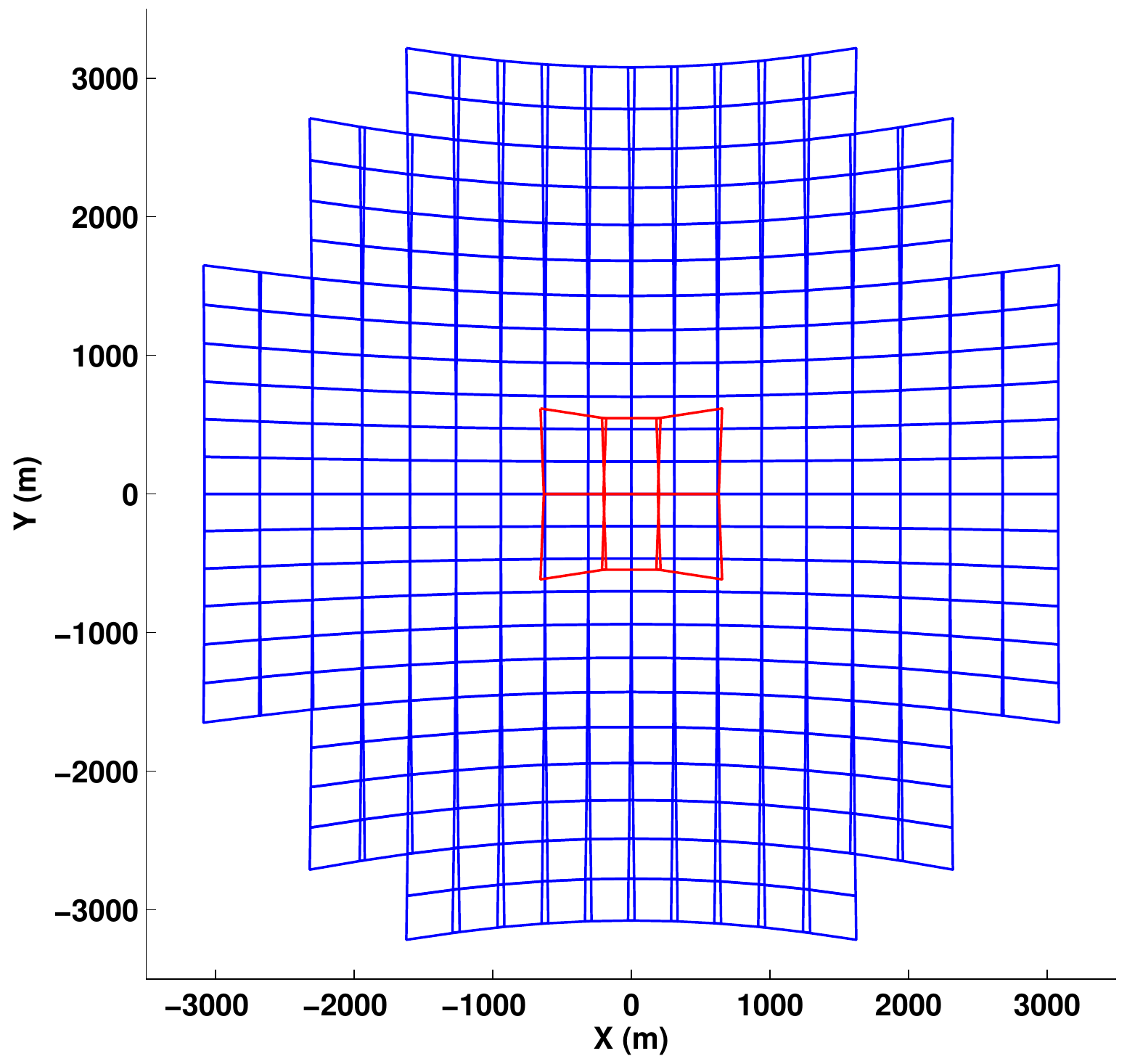}
\caption{The ARGUS-IS uses 368 cameras in a partially populated 24 by 18 array, each camera is 5 megapixels, creating a 1.8 GigaPixel array.  The system typically operates at 17500ft (5333m)~\cite{argus_altitude1,argus_altitude2} creating a 13 cm per pixel ground resolution (at the center of a nadir pointing image, this reduces towards the edge of the sensor).  The sensor footprint at this altitude is shown here together with a comparison (red lines) against the Columbia Large Image Format (CLIF) imager at the equivalent ground resolution.}
\label{fig:argus_array}
\end{figure}

\begin{figure}
\centering
\includegraphics[width=0.8\columnwidth]{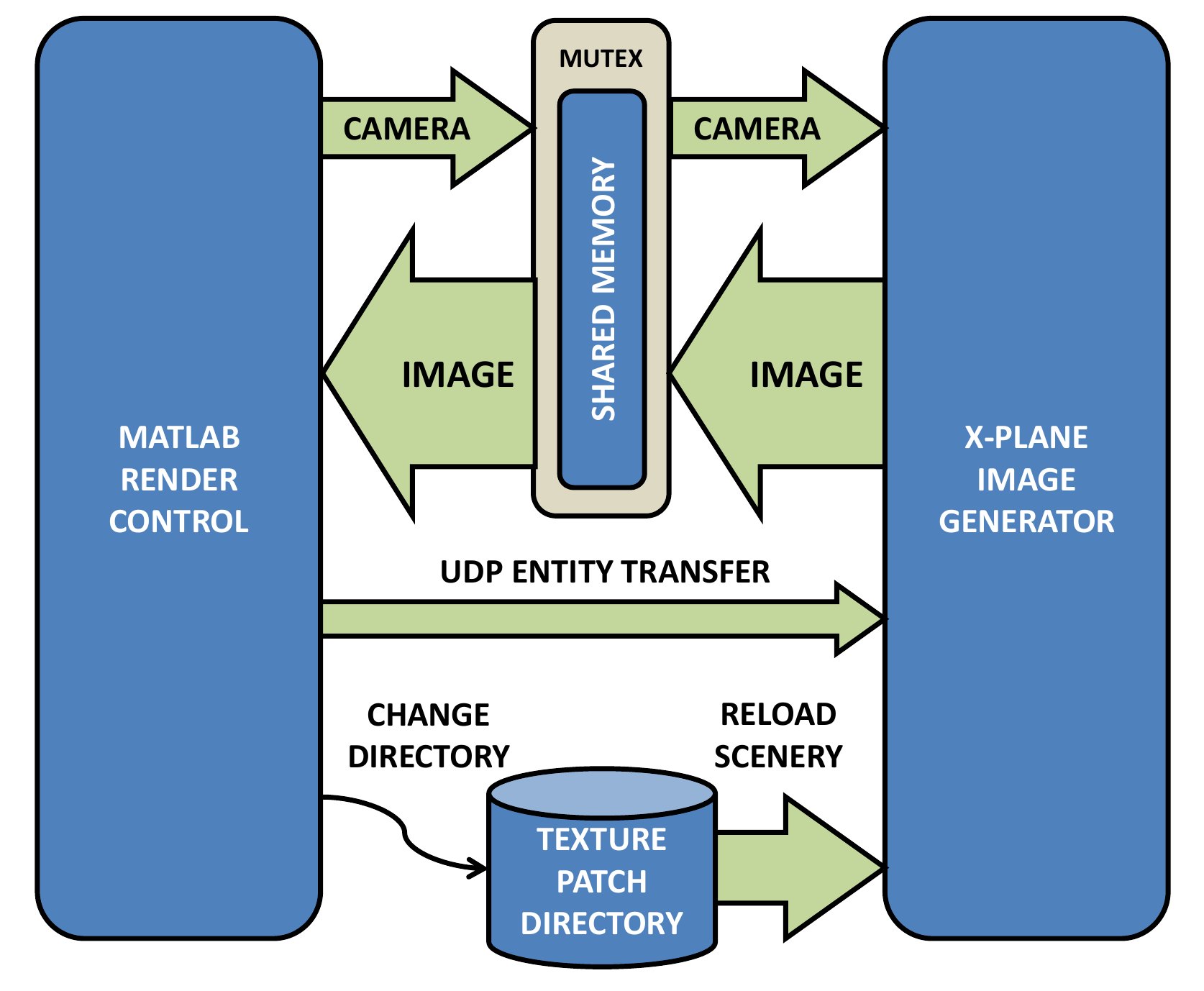}
\caption{The rendering system comprises of an Image Generator (X-Plane 10) controled by MATLAB.  The positions of people and vehicles are transferred via a UDP network connection and rendered.  The camera position is set via a shared memory interface and is followed by a corresponding capture and transfer of imagery back to MATLAB, this interface is protected by a shared mutex that ensures a complete frame is written prior to reading.  As the system has finite resources, very high resolution (15cm per pixel) texture patches are controled externally to provide texture coverage across the wide area (excess of 36km$^2$)}
\label{fig:render_system}
\end{figure}
\begin{figure}
\centering
\includegraphics[width=0.8\columnwidth]{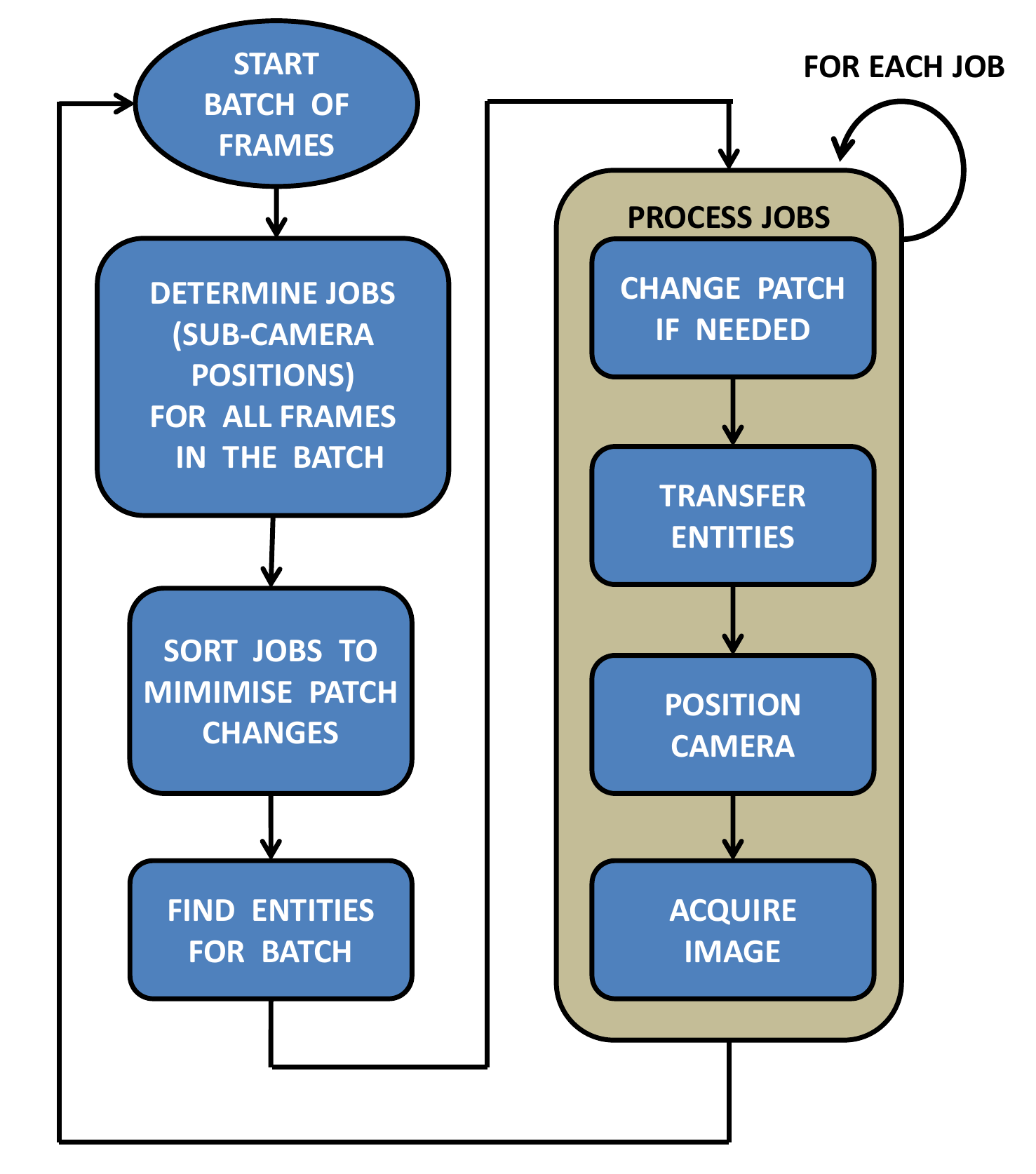}
\caption{The rendering cycle has two parts, batch assembly and job processing.  The camera positions for all the frames in the batch are calculated, and their intersection with the terrain.  This intersection is used to determine which texture patches, pedestrians and vehicles are required.  Each camera considered is a ``job" and these are re-sorted in order to minimize texture patch loading (i.e. grouped by patch ID).  The batch (list of jobs) is passed to the job processor, which changes the texture patch if required, transfers the vehicles and pedestrians (entities) to the Image Generator, positions the cameras, and captures the image.  This process is repeated until the job list is exhausted.}
\label{fig:render_flow}
\end{figure}

\begin{figure}
\centering
\includegraphics[width=\columnwidth]{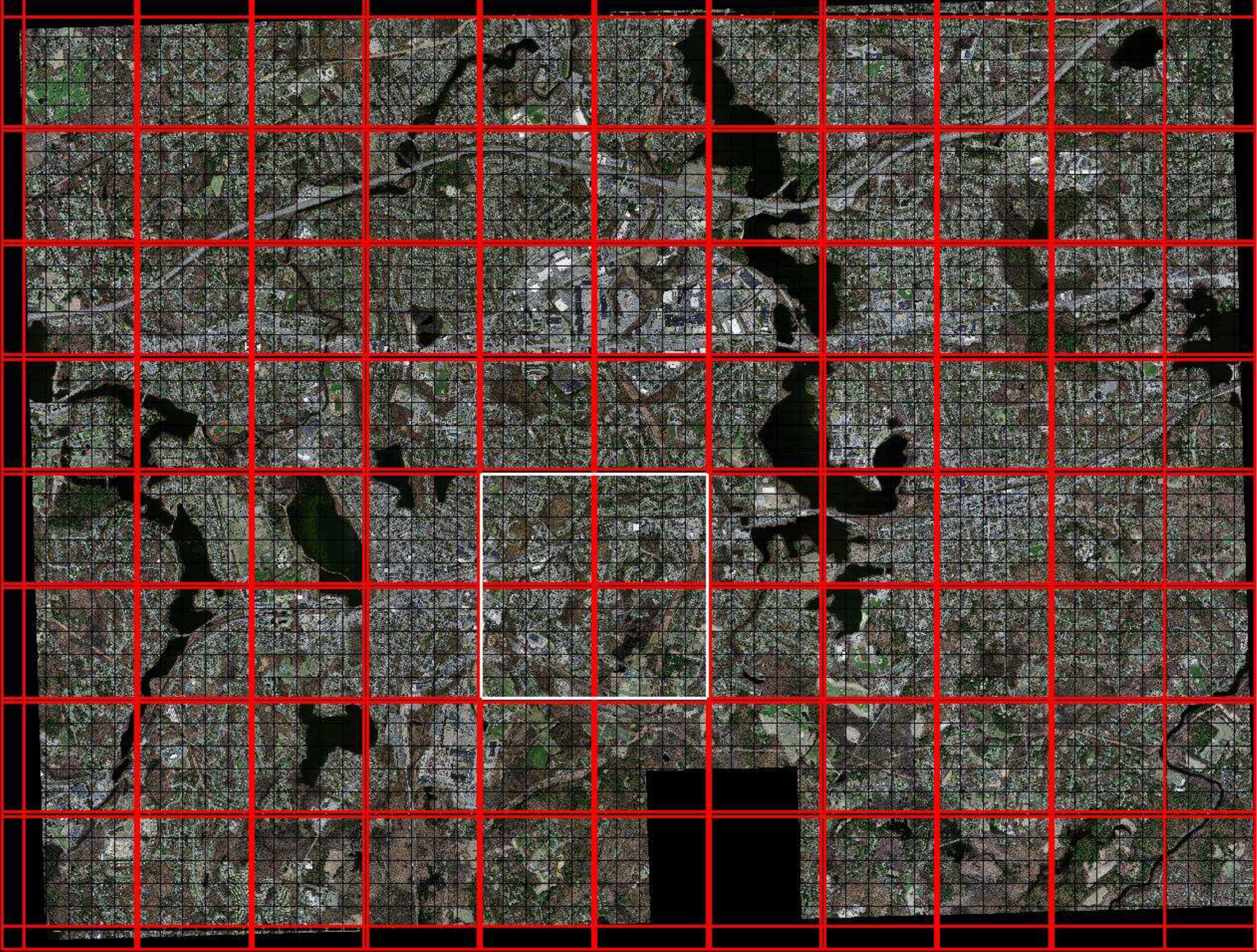}
\caption{This image shows approximately 27 Gigabytes of 15cm per pixel imagery.  So that the Image Generator is not overwhelmed by these large memory requirements, the high resolution textures (2048 by 2048 pixels) are divided into groups of 10 by 10 polygons (red squares) with a 5 polygon overlap - a texture patch.  For the scenarios under investigation, each subcamera footprint will fit inside one of these patches.  When a patch is loaded into the X-Plane Image Generator, all subcameras are rendered (out of sequence) for this patch prior to unloading and loading a new patch.  The black square at the bottom of the figure is an area not captured by the USGS at 15cm, and could be replaced with a different resolution (30cm) image if required.}
\label{fig:texture_division}
\end{figure}

\begin{figure}
\centering
\includegraphics[width=\columnwidth]{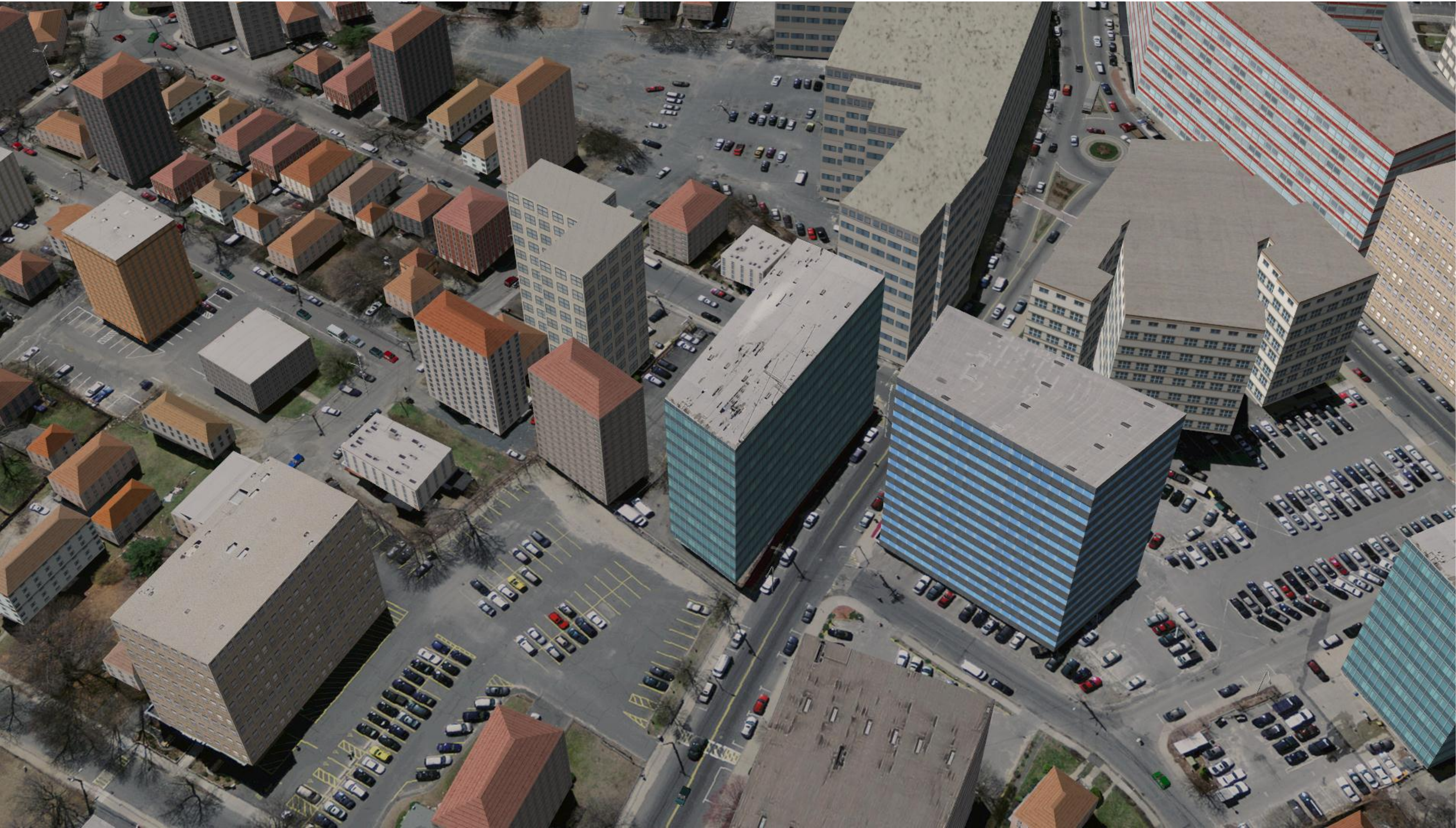}
\caption{Example imagery taken at very low altitude and 45 degree slant angle, showing 3D buildings, vehicles (automobiles and motorcycles) and people.  There are a number of static vehicles present during the capture of the ground texture, and automated processes for detecting and removing these artefacts are  under consideration. An example video is available at: https://stream.liv.ac.uk/zbj9sswg}
\label{fig:low_slant_angle}
\end{figure}

\begin{figure}
\centering
\includegraphics[width=\columnwidth]{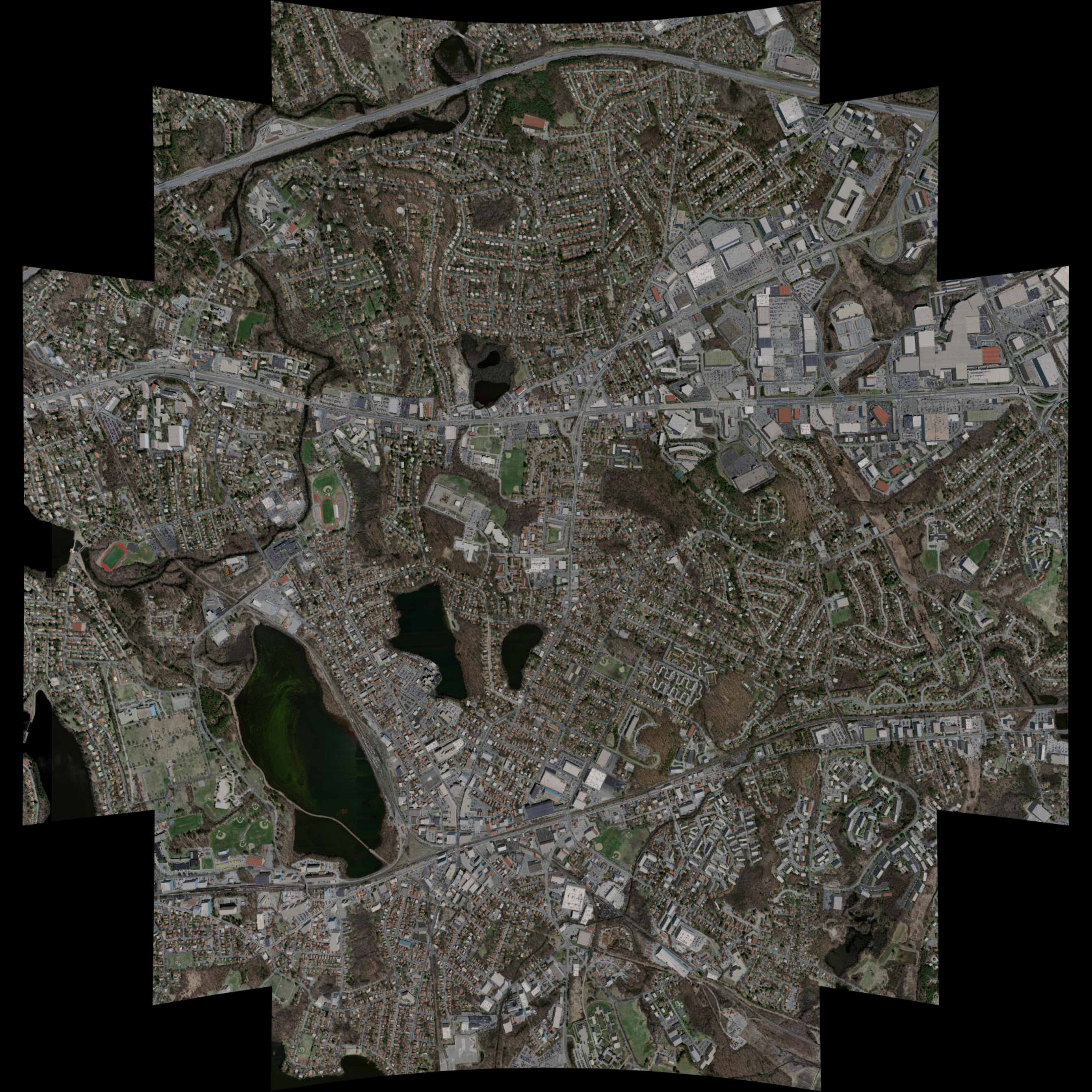}
\caption{The 1.8 Gigapixel ARGUS-IS mosaic, assembled from 368 cameras, this stitched image occupies 5.18 Gigabytes per frame (uncompressed) and must be stored in the BigTIFF file format as it exceeds the 4 Gigabyte limit of the TIFF file format.  Alternatively the stitched mosaic can be subdivided into a tileset that can be quickly concatenated to form the mosaic.  The entire area is simulated at the level of detail shown in Figures~\ref{fig:low_slant_angle} and~\ref{fig:argus_zoomed}. An example video is available at: https://stream.liv.ac.uk/y24dgbey}
\label{fig:argus_mosaic}
\end{figure}

\begin{figure}
\centering
\includegraphics[width=\columnwidth]{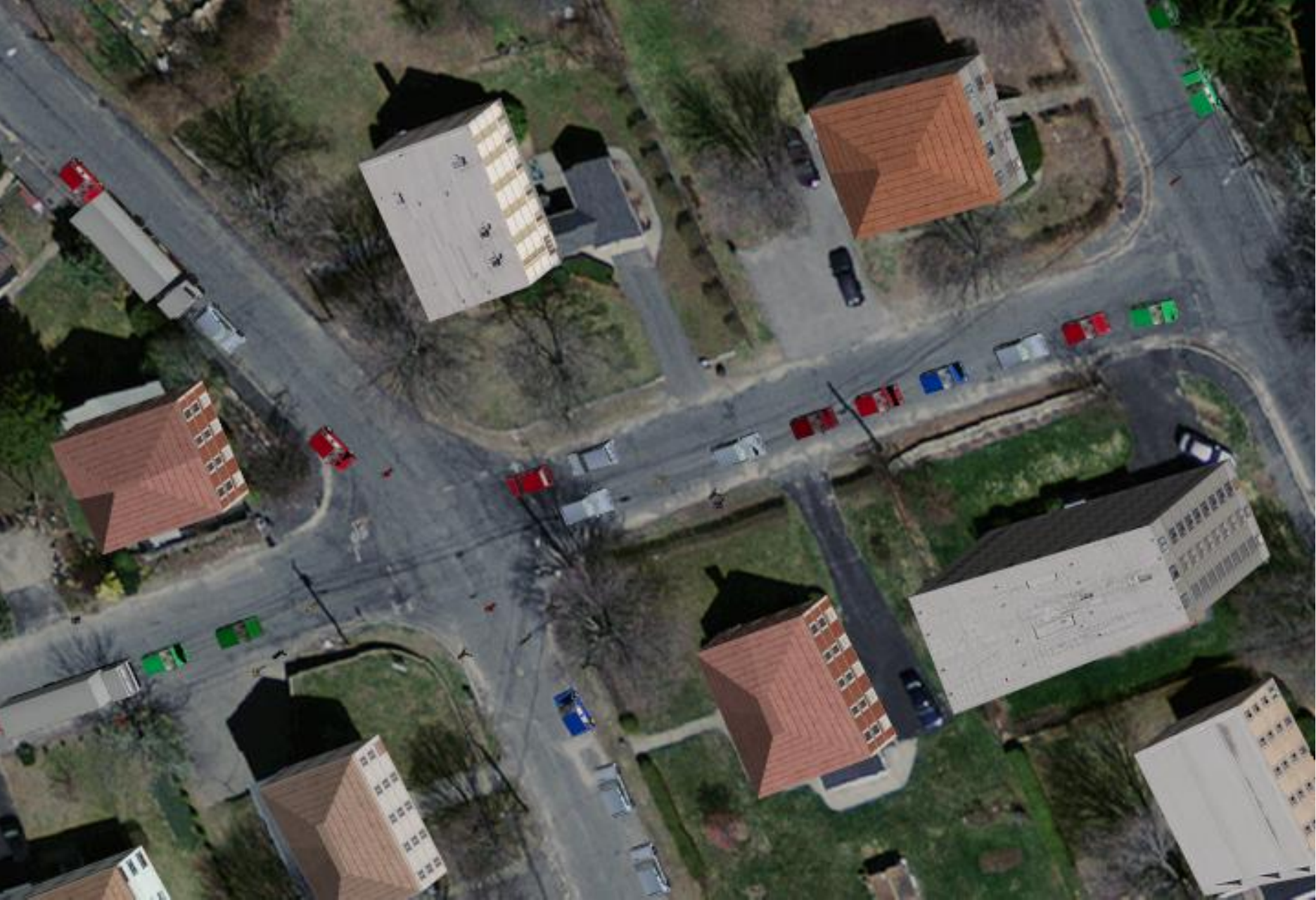}
\caption{Region within the wide area figure~\ref{fig:argus_mosaic}, where the entire area is simulated and captured at this level of detail.  People can be seen crossing the junction whilst the vehicles wait.  This 720 by 490 pixel figure occupies 8.7cm by 5.9cm, whereas if the 41888 by 41888 pixel figure~\ref{fig:argus_mosaic} were to be printed at the same scale, it would measure 5.06 by 5.06 meters.}
\label{fig:argus_zoomed}
\end{figure}

{\section{Wide Area Imager}\label{fig:chapter3}}

\noindent Large scale, wide area, high resolution video is a commonly used form of persistent surveillance data~\cite{menthe2012future}.  However, obtaining such data sources is difficult and requires specialized camera systems with access to aircraft platforms - which may be in demand as these platforms are expensive or military owned, and require significant organization with specialist facilities and personnel (airport, pilot, maintenance) required outside a typical image analysis group.   In addition, there may be issues with obtaining permission for the release of data, not only for military operations but also civilian data that is often under data protection laws~\cite{senior2003blinkering}.  Such aerial video may not be of sufficient resolution to identify faces of people or vehicle models, however the start point and stop point of processed tracks could point to real world building addresses which can then be linked to individuals.

Synthetic video allows rapid modification of the environment and flight path, it also allows extreme and rare event conditions and camera angles to be tested with specific criteria. In addition, imagery generated from a known positional source such as the city model described in Chapter~\ref{fig:chapter2} has the significant advantage of exact and automatically obtained ground truth data.
\begin{table}
\centering
\caption{Imager constants}
\label{table:imager_const}
\begin{tabular}{|c|c|c|c|}
\hline
Symbol & Name & Value & Ref\\
\hline
$z$ & Typical operating altitude & 17500ft (5334m) & \cite{argus_altitude1,argus_altitude2}\\
$FOV$ & Total field of view (circular) & 60 degrees & \cite{argus_datasheet, leininger2008autonomous}\\
$\Delta\theta$ & Angular resolution of 1 pixel & 25 microradians & \cite{leininger2008autonomous}\\
$F_{c}$ & Frame rate (captured) & 10Hz & \cite{argus_datasheet}\\
$F_{s}$ & Frame rate (stored to disk) & 2Hz & \cite{argus_datasheet}\\
$PS_{h}$ & Subcamera resolution (width) & 2592 pixels & \cite{MT9P031}\\
$PS_{v}$ & Subcamera resolution (height) & 1944 pixels & \cite{MT9P031}\\
$N$ & Number of subcameras & 368 cameras & \cite{leininger2008autonomous}\\
$N_{h}$ & Number of subcameras (width) & 18 cameras & \cite{argus_altitude2}\\
$N_{v}$ & Number of subcameras (height) & 24 cameras & \cite{argus_altitude2}\\
\hline
\end{tabular}
\end{table}
Current wide area persistent surveillance cameras are assembled as a array of smaller cameras, as it is difficult to manufacture very large sensors without (excessive) errors.  The two sensor arrays considered here are the six camera CLIF-type 95 megapixel imager (red lines Figure~\ref{fig:argus_array}) using parameters published after the WAFPB-2009 data collection~\cite{cohenour2015camera}, and the 368 camera ARGUS-type 1.8 gigapixel imager (blue lines Figure~\ref{fig:argus_array}) using parameters obtained from publicly available BAE Systems datasheets~\cite{argus_datasheet} and information released to the public in other formats~\cite{argus_altitude1,argus_altitude2}.

With one uncompressed ARGUS-IS image frame being 5.18 gigabytes, the raw uncompressed data rates from the camera array associated with these systems can exceed 500 gigabytes per second (at 10Hz), as this would easily saturate typical network connections and require up to 368 rendering instances per ARGUS-IS, the current system in non-realtime renders to video files that can be retrieved, and reused by 3rd parties many times without the full simulation operating.
\begin{eqnarray}
	azimuth(v,h) &=& \frac{FOV}{N_{h}}~\left(h - \frac{1}{2}\right) - \frac{FOV}{2} \label{eqn:argus_subsensor_azimuth}\\	
	elevation(v,h) &=& \frac{FOV}{N_{v}}~\left(v - \frac{1}{2}\right) - \frac{FOV}{2} \label{eqn:argus_subsensor_elevation}\\
	roll(v,h) &=& 0 \label{eqn:argus_subsensor_roll}
\end{eqnarray}

The pointing angles for the 6 subcameras used by the CLIF sensor can be found in~\cite{cohenour2015camera}, however the pointing angles for the ARGUS-IS sensor have not been published, and so an approximation is presented in Equations~\ref{eqn:argus_subsensor_azimuth},~\ref{eqn:argus_subsensor_elevation} and~\ref{eqn:argus_subsensor_roll}.  These equations provide the azimuth and elevation offsets relative to the boresight for a particular subcamera in the rectangular 24 by 18 array, and are added to the overall rotation into world space.  The vertical index $v$ ranges from 1 to $N_{v}$ (the vertical size of the array) where $N_{v}=24$.  The horizontal index ranges from 1 to $N_{h}$ (the horizontal size of the array) where $N_{h}=18$.  As the ARGUS-IS has a circular lens, not all the corner indices are utilized as the array, see Figure~\ref{fig:argus_array} to determine which subcameras are present.  Due to this circular lens, the horizontal and vertical fields of view are both $FOV=60.0$ degrees~\cite{argus_datasheet, leininger2008autonomous}.

{\subsection{Implementation of Capture System}}

\noindent The system described by Figures~\ref{fig:render_system} and ~\ref{fig:render_flow} uses a stop motion animation system to capture video frames in a temporally consistent manner across all the subcameras in the array.  The rendering system uses a commercial off the shelf (COTS) flight simulator, X-Plane 10, as the image generator (IG) controled via a combination of shared memory and network UDP interfaces, as flight simulators are optimized to render large outdoor areas from a flying platform.  

For the WAAS use-case, the texture requirements are both high in resolution (15cm per pixel) and span a visible area of $30-40km^2$ at any one time (dependent on platform altitude) - this does not include the areas that exist outside the camera view but could quickly become visible by turning the sensor.  However, limited texture memory on the GPU restricts the amount of texture data that can be loaded at any one time.  

To attempt to solve this problem and present a pleasant user experience, flight simulators use a bespoke form of Level Of Detail(LOD)~\cite{de1995levels} scaling to maintain a balance between the rendering of the visible area and image quality.  The system described overrides this behavior, to favor texture quality, by externally managing the resources that are available for loading.  When required, the capture controller swaps texture folders and triggers a scenery reload, such that a number of small areas (patches) of very high quality are loaded and unloaded.  The smaller viewport of the image generator cycles through the subcameras in the WAAS array capturing images and swapping textured areas when required, and during this process it is of vital importance to maintain the state of every entity in the scene in order to create a consistent wider image frame.

The City Simulator output is stored in temporally separated XML files, one file per timestep to avoid multi-gigabyte XML files.  Vehicle and pedestrians timesteps are loaded from file into memory for the batch of frames to be rendered.  The visual model type and positional data required to draw these entities are transferred to the X-plane flight simulator via a custom ``plug-in" that receives UDP packets, frustum culling is used to transfer only the entities that are in view which increases the overall rendering performance.

Each captured image is transferred out of the flight simulator and directly into a MATLAB matrix, via a shared memory interface, where memory access is controled by a named mutex to ensure a complete frame is present on each capture.  Shared memory is used for bulk transfer of large image matrices, as the transfer involves only a \textit{memcopy} operation and triggering of a shared mutex.  Camera control is achieved through the same interface to ensure the camera is in place prior to (optional) capture.  These captured image files are saved in the TIF file format to a solid state disk for speed, the CPU processing delay when compressing and uncompressing a file format such as PNG is a significant cost.  The files are compressed in parallel processes, only after the frame batch has been fully processed.

\subsubsection{Data pre-processing}

\noindent Typical image generation focuses scenery modeling effort towards detailing a small area, however an important consideration for WAAS-scale video generation is consistency of scene quality across the whole imager, mixtures of texture quality within the same image is immediately apparent and affect further image processing.  In addition, Open Street Map data must also contain building outlines for autogeneration of three dimensional objects.    

The United States Geological Survey (USGS) provides a large repository of aerial photography~\cite{usgs}, and 1 meter or finer imagery is available for much of the United States, although Massachusetts was chosen for the consistent and near complete presence of building outlines in the Open Street Map data and 30cm coverage of the entire state with 15cm coverage over particular cities.  
The ARGUS-IS has an estimated angular resolution of approximately 25 microradians per pixel~\cite{leininger2008autonomous}, which corresponds to a ground resolution of 13cm at the centre of a nadir pointing camera at 5334m (17500ft)~\cite{argus_altitude1,argus_altitude2}, therefore using 15cm USGS photography as ground textures should yield imagery somewhat similar to the actual sensor.  The georeferenced USGS images are processed into a continuous set of 256 by 256 pixel tiles using the Geospatial Data Abstraction Library (GDAL)~\cite{gdal}, the tiles are then assembled into 2048 by 2048 pixel images and combined with corner locations during conversion to XPlane textured polygons.

Figure~\ref{fig:texture_division} shows approximately 27 Gigabytes of 15cm imagery, each of the grid squares (black squares) is a 2048 by 2048 pixel textured polygon, which are grouped into patches (red squares) consisting of 10 by 10 polygons with a 5 polygon overlap that becomes the area that is loaded into the XPlane flight simulator (white square).  The patches are overlapping and are selected to fully contain the textured area as observed through a single subcamera in the array, each subcamera rendering task is tagged with the patch index required for the scene, and the capture tasks for the array are grouped by this patch index in order to minimize patch swapping (texture unload and reload).  The frames are processed in batches, and therefore the patch index grouping is also extended temporally to process all imagery that requires a texture patch before loading a new one, reducing swapping when compared with resetting at each frame.

Figure~\ref{fig:low_slant_angle} is an image taken from low altitude and shallow slant angle, which shows the three dimensional ground and building detail possible using the X-Plane 10 simulator and USGS textures.  There are 3D-vehicles and people present, however the textures also show vehicles present during the image capture, these will remain static and can be filtered out, but future efforts will cleanse the roads of vehicle images through vehicle detection and replacement through texture duplication~\cite{he2012statistics}.  

\subsection{Data post-processing}

\noindent The imagery has a variety of use-cases which include processing of the raw subcamera images, the whole wide area frame or video.  Acting directly on captured subcamera images requires no further processing of the imagery, however full frame wide area imagery and video requires warping and stitching of the subcamera images to create a large flattened mosaic. 
 
When creating the mosaic, having a full frame in memory is unnecessary, and instead the subcamera images are locally stitched and re-cut into smaller tiles - this allows multiple stitching processes to run in parallel, each working on a separate frame without exceeding memory constraints.   When the full mosaic is required, it is assembled from these tiles with simple concatenation.  Groups of these tiles can also be loaded into video windows and played back sequentially. 

The subcamera corners in the mosaic were located by projecting the subcamera corners on to a flat image plane, resembling Figure~\ref{fig:argus_array}.  Equation~\ref{eq:separation_distance} defines the plane-eyepoint separation distance $R$ which is selected to create a pixel coordinate system where the central area of the mosaic is of equal resolution to the original subcamera images, so the image quality is better preserved under image warping.

The equation for $R$ is:
\begin{equation}
	R = \frac{1}{\theta}
	\label{eq:separation_distance}
\end{equation}
\noindent where $\theta$ is the arc subtended by 1 pixel, in the case of the ARGUS-IS sensor $\theta$ is approximately 25 microradians, thus $R\approx 41888$ pixels. (Fig.~\ref{fig:argus_mosaic})

Figures~\ref{fig:argus_zoomed} shows the ARGUS-IS ground level detail which compares very favorably with the published image in~\cite{argus_datasheet}. The resolution window shown in figure~\ref{fig:argus_zoomed} is 720 pixels by 490 pixels and occupies 8.7cm by 5.9cm\textbf{}, whereas the full field of view mosaic spans 41888 by 41888 pixels, therefore if figure~\ref{fig:argus_mosaic} were to be printed at the same scale as figure~\ref{fig:argus_zoomed} (on A4 paper) it would be equivalent to X metres by Y metres.

\begin{figure}
\centering
\includegraphics[width=0.5\columnwidth]{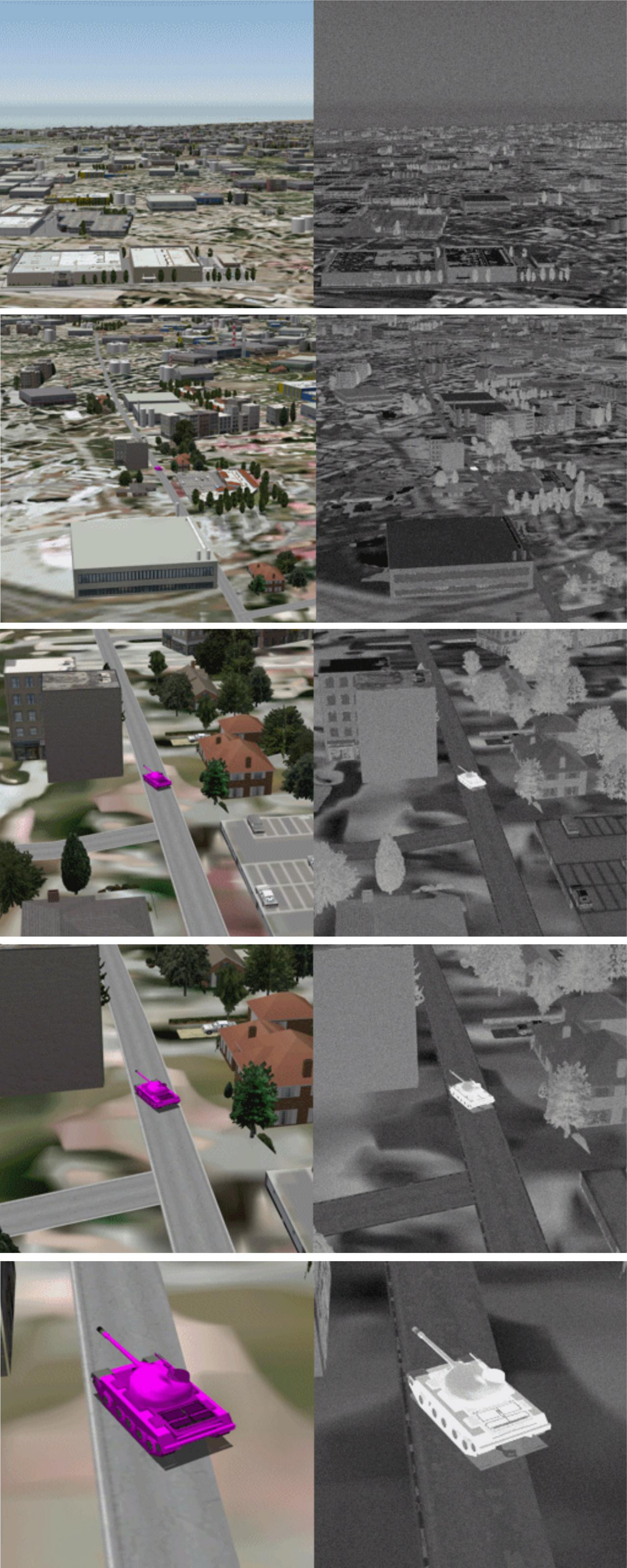}
\caption{Sequence of images, showing an approach to a target (tank).  The 1st column is the RGB source imagery with the thermal target colored magenta to separate it from the random background clutter.  The 2nd column shows the transformed infrared imagery, with noise sources and non-uniformities added.}
\label{fig:infrared}
\end{figure}

\begin{figure}
\centering
\includegraphics[width=0.75\columnwidth]{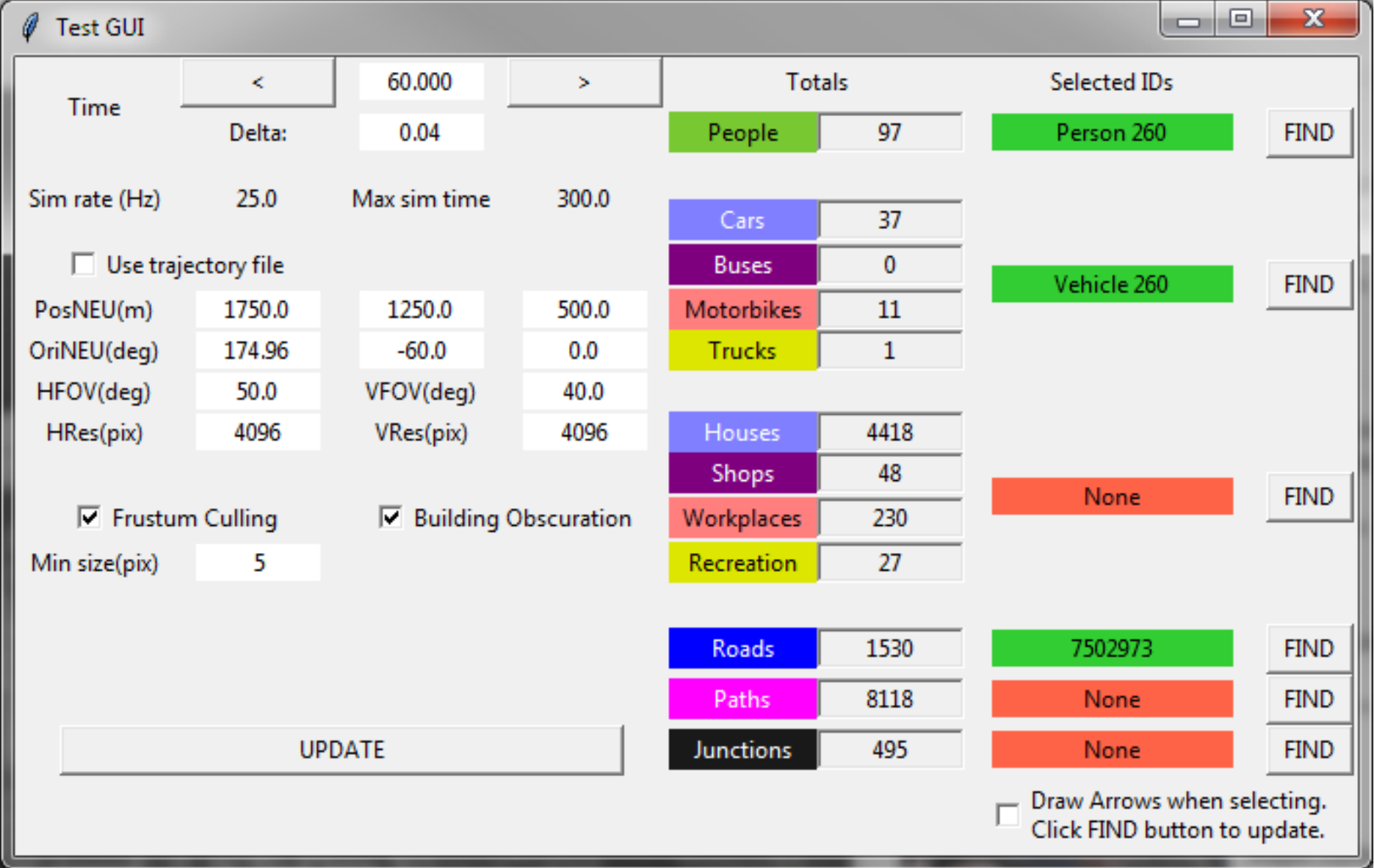}
\\\vspace{5pt}
\includegraphics[width=0.75\columnwidth]{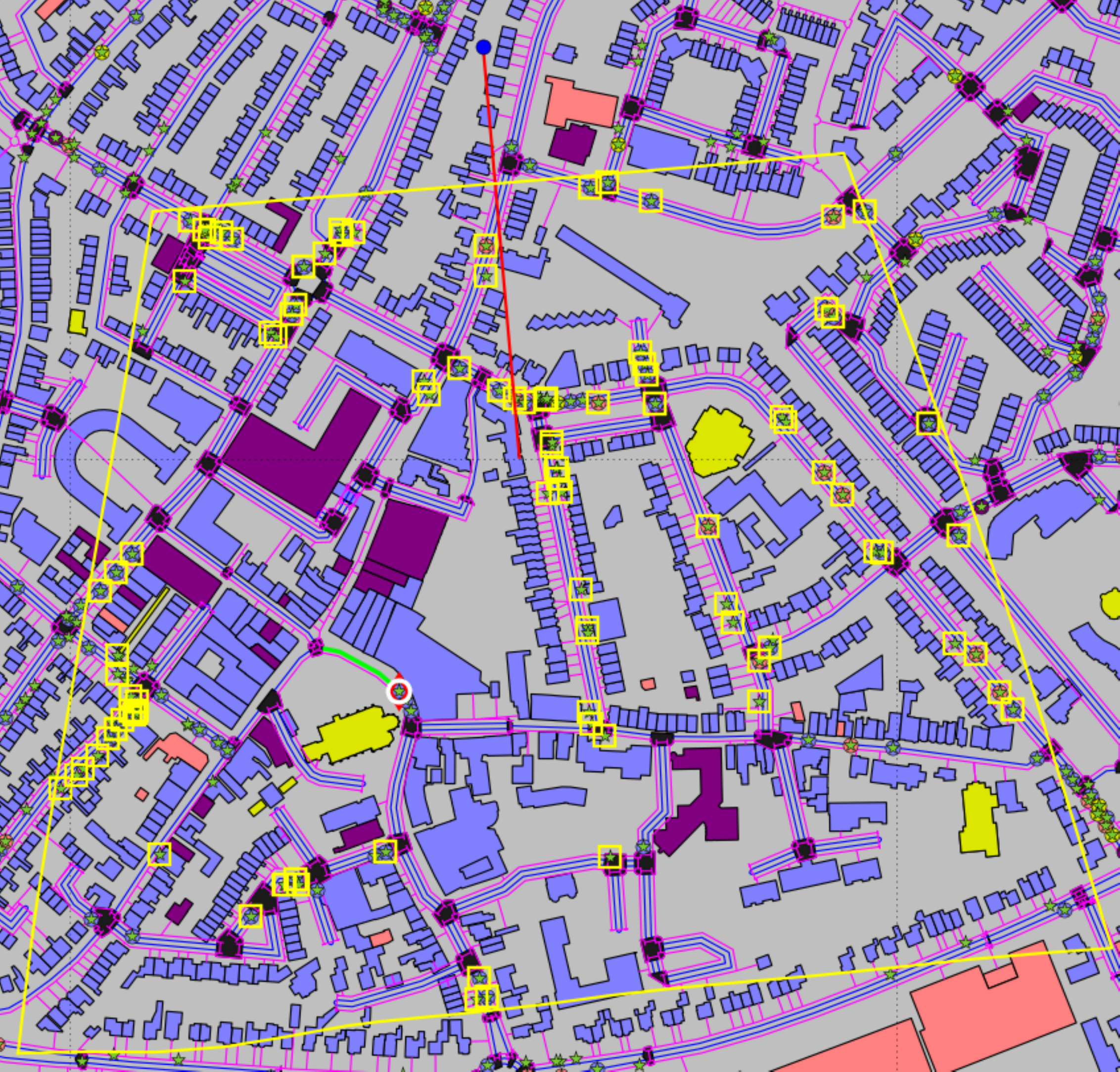}
\caption{The Graphical User Interface (GUI) for exploring the ground truth data produced by the city simulation, also allows identification and location of objects by ID (Identification string).  In addition to visualizing the data as a map, the GUI uses a virtual camera to determine visibility of objects independent of camera resolution.  This also allows a precise count of the true number of visible objects, without the need for image generation and processing such as the separation of individual objects in an image.}
\label{fig:gui}
\end{figure}

\begin{figure}
\centering
\includegraphics[width=\columnwidth]{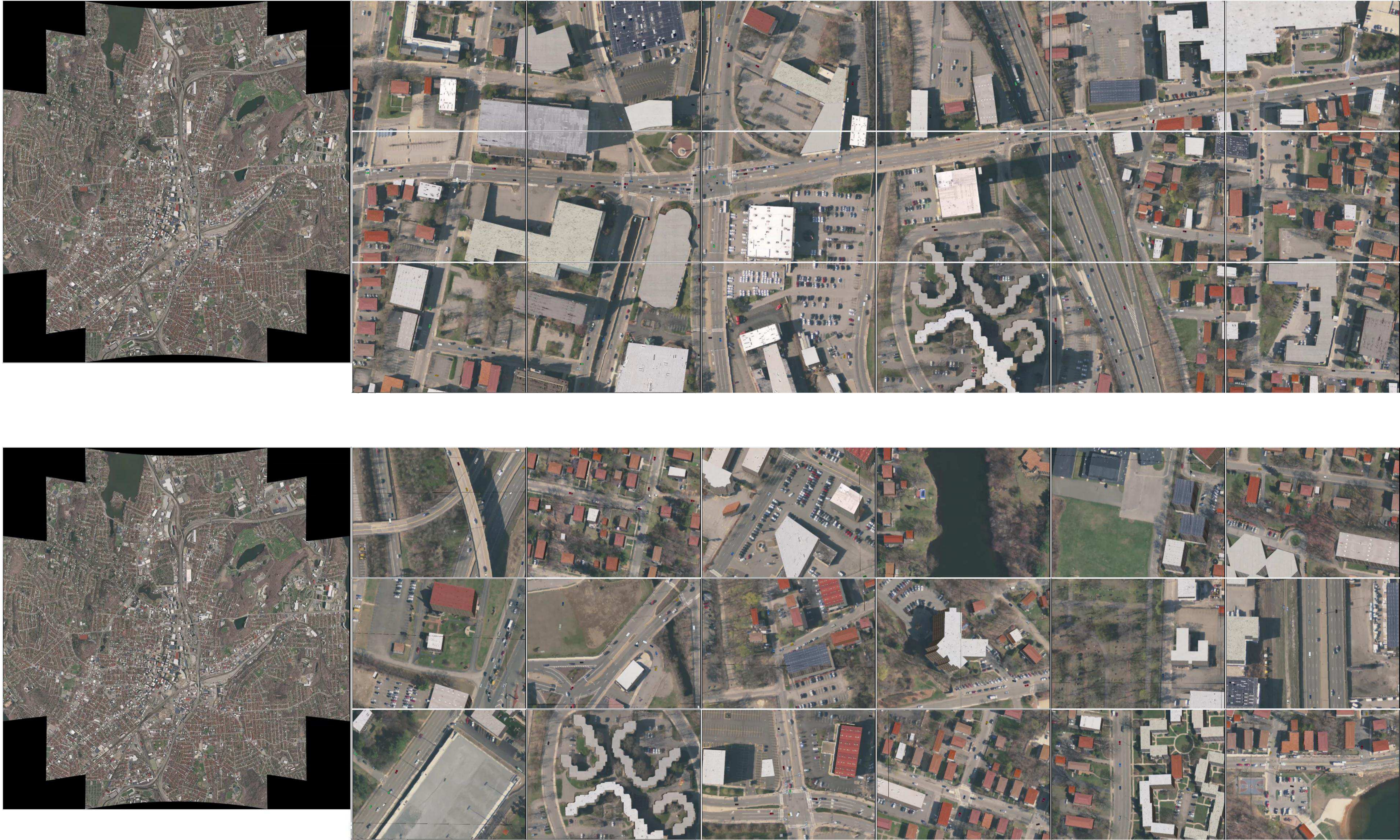}
\caption{The WAAS video playback is performed through a specialized tool with multiple viewing windows.  The overview is also a video, and subwindows can be selected interactively from within the overview to either form a contiguous mosaic \textbf{(top)} or viewpoints from anywhere on the overview \textbf{(bottom)}.}
\label{fig:argus_viewer}
\end{figure}

\subsection{Infrared processing}

\noindent The system is not constrained to only modeling visible band imagery. It can be modified to generate images corresponding to thermal infrared bands through the use of specific color channels. Infrared imagery has three-dimensional structural content that is similar to the visible band, but the intensity depends mainly on the temperature of objects rather than their reflectivity and color. To represent intensity variations, the thermal properties in a scene can be encoded in terms of colors; a particular color represents a particular temperature. To do this for a whole scene, and for the very large scenes presented in this paper, would be extremely time-consuming. As a result, a pragmatic approach is adopted here. The majority of the background scene is allocated a temperature determined by the intensity and the color of the visible band scene (after removing the most obvious effects present in the visible band, such as shadows). Specific objects of interest are then colorized using a special palette of colors which is used to represent more sophisticated thermal textures. In the work presented here, the special palette uses the magenta range of colors, which rarely occur in natural visible band scenes and can easily be separated from the background. A complex object with known thermal properties can be colored using different shades of magenta to represent the different temperatures present, e.g. a car with a hot engine can be colored to provide a hot hood and exhaust. To avoid problems with existing areas of magenta present in the visible band background image, the background is scanned and the hue values of any background regions containing magenta are adjusted to remove the possibility of unwanted artefacts being created in this way.

Using this combination of temperature maps, background and special palette, the expected photon flux produced by each area of the scene is calculated for the appropriate waveband and then an image is generated that is representative of that thermal band. To the photon flux, the properties of the atmosphere can then be added, including wavelength dependent attenuation and atmospheric path radiance, both of which can have a significant effect on the quality of an infrared image. Attenuation is due to absorption of photons by the atmosphere, which depends on range, and path radiance arises because the atmosphere itself has a temperature and emits some thermal radiation, which tends to suppress contrast for objects at longer ranges. As such, both of these processes require a range map or range ‘image’ to be generated (using geometric ray tracing) to allow the inclusion their effect in a physically realistic manner. After the atmospheric properties have been included, an infrared sensor model is used to generate the infrared image; this typically includes~\cite{driggers2012introduction,mooney1996characterizing}: an optical distortion, a point spread function, vignetting, the finite size of the photo-detectors in each pixel on the focal plane array, the efficiency of the photo-detectors, the gain-offset properties of the focal plane array, and the quantization of the signal from the analog-to-digital converters.

{\section{Querying Tool and WAAS Video Playback}\label{fig:chapter4}}

\noindent Generating imagery from a known dataset, has the significant inherent advantage of perfect ground truth positional data which allows comparison of image processing algorithm outputs against exact known values (ground truth).  In addition, non-image processing applications can sample the perfect dataset without the need for complex image processing to obtain a subset of positional data, for example tracks of particular vehicles hidden by buildings.  The ground truth data also has metadata associated with it, for example vehicle type, person name, residence and place of work, with which every data point can be associated with, as unbroken tracks.  To explore this data, an application has been created that allows querying of metadata, locating entities interactively and viewing positional data.  The tool also performs visibility checks on the entities using ray intersections with the scene geometry, this allows fast and precise determination of individual entity visibility independent of the camera resolution and without the need to render and analyze the scene.

Figure~\ref{fig:gui} shows the GUI and map components of the tool displaying the City Simulator output, with timestep controls (top-left), camera controls (left), visible totals (middle) and interactive search functionality (right).  The camera controls not only interactively set a position and orientation, but also allow investigation of camera resolution (via minimum detectable target size) and occlusion by frustum culling and building obscuration.  The map shows color coded buildings, vehicles (circles) and people (green stars) with a camera with bore-sight vector present (blue circle and red line).  The selected person (red diamond) and vehicle (white circle) can be seen on a selected road segment (green) with the identifiers of this selection shown in the GUI text boxes.  The camera viewing area is defined by the yellow line, the curved shape of the line shows the terrain is not flat, and with the frustum culling and building obscuration options active, yellow boxes are used to highlight entities visible to the camera.  

The visibility is calculated using a Ray-Triangle intersection algorithm acting on terrain and building geometry that has be decomposed into triangle meshes.  The terrain is divided into mesh patches and each building is also a separate mesh, every mesh is surrounded by a precalculated Axis Aligned Bounding Box (AABB)~\cite{williams2005efficient} which is used for very efficient downselection of geometry prior to Ray-Triangle intersection.  The selected intersection algorithm by Sunday~\cite{dan_sunday} employs precalculated normals, cross and dot products, avoiding the need to calculate costly cross products on each triangle test.  For static scenes, such as buildings in a city, the Sunday algorithm has a computational advantage compared with the popular M\"{o}ller-Trumbore algorithm~\cite{moller2005fast} that requires two cross products to be calculated for each triangle test.

\subsection{Video playback tool}

\noindent To visualize generated video sequences, a viewing tool was created that allows selection of video windows across the wide field of view, and these windows can be selected at random or mosaic-ed together such that vehicles pass between videos (Figure~\ref{fig:argus_viewer}).  The viewing tool was implemented in C++ using the \textit{libVLC} interface of the VLC media player, and consists of 18 viewing windows and one video overview spread across two 1080p monitors.

{\section{Conclusions} \label{fig:chapter5}}

\noindent This paper has presented a multi-faceted simulation of complex urban environments, which can be used to generate large-scale image and video datasets for the development of algorithms for surveillance and reconnaissance applications. The work has been motivated by the development of wide field of view multi-camera sensors for airborne platforms. These systems, referred to as Wide Area Aerial Surveillance (WAAS) systems, allow large areas of the ground to be monitored for long periods of time and generate very large amounts of image data. They are time-consuming and costly to operate, which means that there are relatively few datasets available for algorithm development. The system presented in this paper includes all of the main aspects necessary for the generation of simulated video datasets, including representative vehicle motion and traffic flows, and pattern of life information. To achieve this, the simulated WAAS system brings together aspects of the urban planning, image generation, Big Data and target tracking research communities. The video data produced using this system is able to cover areas of the ground consistent with the largest of the WAAS sensors, circa 40 km$^2$ for the ARGUS-IS sensor, with traffic flows and pattern of life showing the motion of up to 150,000 individual people and vehicles.

The system makes extensive use of existing software, the use of the X-Plane flight simulator provides an Image Generator that increases in image fidelity as new versions of the software becomes available, and take advantage of new graphics hardware and features. Similarly, the use of SUMO allows new driving models, microsimuation details and features to be automatically integrated into the system, and access to a traceable open source peer-reviewed model.  Due to this, the effort can be focused on data preparation, with moves towards automated cleansing of aerial photography to remove vehicles and other transient artefacts, and the automated planting of 3D trees on top of 2D tree images.

The future direction of the City Simulator will aim to perform large scale distributed rendering of imagery using supercomputers, and the addition a number of new sub-simulation components: Vehicle parking simulations (both large and multistory carparks, plus individual driveways), so that the vehicles are persistent in the simulation when they are idle, and people walk to the parked vehicle to start driving - navigating the carpark in vehicle and importantly also on foot; Pedestrian crowd simulation to simulate complex mixing behavior in open regions, and natural waiting and boarding behaviors at bus stops.  Using the existing framework, these features would be easily integrated as vehicles and people are transferred amongst subsystems via the City Controller in the same manner as the existing SUMO and building footpath sub-simulators.  

{\section{Acknowledgements}}

The authors would like to thank Dawne Deaver and David Oxford from the PerSEval project for helpful and informative discussions in the early stages of this work. This work was supported by the Defence Science and Technology Laboratory, under a subcontract from Roke Manor Research Ltd (Chemring Group), on behalf of the UK Ministry of Defence.

{\section{Bibliography}}

\bibliography{WAMI_paper}

\end{document}